\documentclass{article}
\usepackage{graphicx}
\usepackage{amssymb}
\usepackage{amsmath}
\usepackage{epstopdf}
\usepackage{authblk}
\usepackage{tikz}
\usepackage{booktabs}
\usetikzlibrary {arrows.meta}
\usepackage{amsthm}
\usepackage{subcaption}
\usepackage{multirow}
\usepackage{float}

\usepackage[all]{xy}
\usepackage[plain]{algorithm2e}
\usepackage{colortbl}
\newcommand{\T}{\mathrm{T}}

\newcommand{\Z}{\mathrm{true}}
\title{TRENDY: Gene Regulatory Network Inference Enhanced by Transformer}
\author[1]{Xueying Tian}
\author[2]{Yash Patel}
\author[3,*]{Yue Wang}

\affil[1]{School of Information, University of California, Berkeley, CA 94720, USA}
\affil[2]{Department of Mathematics, University of Miami, Coral Gables, FL 33146, USA}
\affil[3]{Irving Institute for Cancer Dynamics and Department of Statistics, Columbia University, New York, NY 10027, USA}
\affil[*]{Corresponding author: yw4241@columbia.edu}
\date{}

\begin{document}

\maketitle

\begin{abstract}
Gene regulatory networks (GRNs) play a crucial role in the control of cellular functions. Numerous methods have been developed to infer GRNs from gene expression data, including mechanism-based approaches, information-based approaches, and more recent deep learning techniques, the last of which often overlook the underlying gene expression mechanisms. In this work, we introduce TRENDY, a novel GRN inference method that integrates transformer models to enhance the mechanism-based WENDY approach. Through testing on both simulated and experimental datasets, TRENDY demonstrates superior performance compared to existing methods. Furthermore, we apply this transformer-based approach to three additional inference methods, showcasing its broad potential to enhance GRN inference.
\end{abstract}

\section{Introduction}
The expression of genes can be regulated by other genes. We can construct a directed graph, where vertices are genes, and edges are regulatory relationships. This graph is called a gene regulatory network (GRN). GRNs are essential to understanding the mechanisms governing cellular processes, including how cells respond to various stimuli, differentiate, and maintain homeostasis \cite{wang2022chronic,wang2020identification}. Understanding the GRN structure is important for developmental biology \cite{wang2020biological,cheng2024reconstruction,cheng2022ex,sha2024reconstructing,myasnikova2020gene}, disease research \cite{mcdonald2023computational,cheng2023mathematical,angelini2022model,arshad2014using}, and even studying macroscopic behavior \cite{li2021chronic,vijayan2022internal,axelrod2023drosophila}. 

It is difficult to directly determine the GRN structure with experiments. Instead, researchers develop methods that infer the GRN structure through gene expression data. Some methods, such as WENDY \cite{wang2024gene} and NonlinearODEs \cite{ma2020inference}, are mechanism-based: they construct dynamical models for gene expression, and fit with expression data to determine the GRN \cite{burdziak2023sckinetics,wang2023dictys}. Some methods, like GENIE3 \cite{huynh2010inferring} and SINCERITIES \cite{papili2018sincerities}, are information-based: they treat this as a feature selection problem and directly find genes that { can be used to predict the level of the target gene (e.g., by linear regression). Then this predictability (information) for the target gene is assumed to imply regulation} \cite{zheng2019bixgboost,huynh2018dyngenie3}. 

Recently, there are some deep learning-based methods that infer GRN with different neural network frameworks, such as convolutional neural networks \cite{nauta2019causal}, recurrent neural networks \cite{kentzoglanakis2011swarm}, variational autoencoders \cite{shu2021modeling}, graph neural networks \cite{feng2023gene,mao2023predicting}, etc. Deep learning frameworks can perform well on many tasks. However, because of the large number of tunable parameters in a neural network, it is difficult to explain why it works. These deep {learning-based} GRN inference methods have similar problems: they generally apply popular neural networks as black boxes without integration with biological mechanisms, and thus lacking interpretability.

The transformer model is a deep learning architecture designed to handle sequential data \cite{vaswani2017attention}. It can capture long-range dependencies and relationships {(linear and nonlinear)} through the self-attention mechanism. The encoder layer of the transformer model can process the input through a neural network, so that the output is close to the given target. 

Some researchers have developed GRN inference methods based on the transformer model \cite{xu2023stgrns} or its variants \cite{shu2022boosting}. The STGRNS method \cite{xu2023stgrns} predicts the existence of regulation between two genes solely based on the expression levels of these two genes. Since other genes are not considered, this approach makes it difficult to distinguish between direct regulation ($G_i\to G_j$) and indirect regulation through a third gene ($G_i\to G_k\to G_j$). The GRN-transformer method \cite{shu2022boosting} uses multiple inputs, including the GRN inferred by an information-based method, PIDC \cite{chan2017gene}, but it also does not consider the biological dynamics of gene regulation.

To combine the powerful deep learning techniques and biological understanding of gene regulation, we propose a novel GRN inference method that applies transformer models to enhance a mechanism-based GRN inference method, WENDY. This new method, TRansformer-Enhanced weNDY, is abbreviated as TRENDY. {WENDY method is based on a dynamical model for gene regulation, and an equation for the GRN and the covariance matrix of genes is solved to derive the GRN}. In TRENDY, we first use a transformer model to construct a pseudo-covariance matrix that performs better in WENDY. Then we apply another transformer model that directly enhances the inferred GRN.

The idea for the second half of TRENDY can be used to enhance the inferred results by other GRN inference methods. We apply transformer models to three GRN inference methods, GENIE3, SINCERITIES, and NonlinearODEs, to obtain their {transformer-enhanced versions, tGENIE3, tSINCERITIES, and tNonlinearODEs.}

{There have been some research papers that study the idea of denoising networks (e.g., enhancing GRNs inferred by other methods), such as network deconvolution (ND) \cite{feizi2013network}, diffusion state distance \cite{cao2013going}, BRANE Cut (BC) \cite{pirayre2015brane}, BRANE Clust \cite{pirayre2017brane}, and network enhancement \cite{wang2018network}. }

{We test the four traditional methods (WENDY, GENIE3, SINCERITIES, and NonlinearODEs) with their enhanced versions (by transformer, ND, or BC) on two simulated data sets and two experimental data sets. All four transformer-enhanced methods outperform other methods, and TRENDY ranks the first among all methods.}

In order to train a deep learning model, such as the transformer model used in this paper, we need a sufficient amount of training data. Specifically, since we want the GRN inference method to work well in different situations, the training data should contain many different data sets that correspond to different GRNs. In reality, there are not many available experimental gene expression data sets with known GRNs. Therefore, we consider generating synthetic data sets. 

To simulate new gene expression data, the most common approach is to assume that gene expression follows a certain mechanism, such as an ordinary differential equation (ODE) system \cite{hache2009genge,kamimoto2023dissecting} or a stochastic differential equation (SDE) system \cite{dibaeinia2020sergio,schaffter2011genenetweaver}. After determining the parameters of this system, one can simulate a discretized version of this system and obtain new data. Readers may refer to a review \cite{tripathi2017sgnesr} for this mechanism approach. Another approach is to apply deep learning to learn from experimental gene expression data and generate new data that mimic the existing data. Such generative neural networks can be variational autoencoders \cite{shu2021modeling} or generative adversarial networks \cite{marouf2020realistic}.

The advantage of the differential equation approach is that one can change the parameters to generate new data that correspond to another GRN, while the deep learning approach can only reproduce existing data. The disadvantage of the differential equation approach is that the modeled mechanism might not be a perfect fit to reality. Therefore, the generated data might be questionable, especially those generated by an oversimplified model, such as a linear ODE system. Instead, the deep learning approach can generate new data that are indistinguishable from the experimental data. 

In this paper, we need to feed the transformer models with data from many different GRNs. Therefore, we adopt the mechanism approach with a frequently used nonlinear and stochastic system \cite{pinna2010knockouts,papili2018sincerities,wang2024gene}:
\begin{equation}
\mathrm{d}X_j(t)=V\left\{\beta \prod_{i=1}^n \left[1+(A_\Z)_{i,j}\frac{X_i(t)}{X_i(t)+1}\right]-\theta X_j(t)\right\}\mathrm{d} t +\sigma X_j(t)\mathrm{d}W_j(t).
\label{eqnl}
\end{equation}
The matrix $A_\Z$ is a randomly generated ground truth GRN, where $(A_\Z)_{i,j}>0/=0/<0$ means that gene $i$ has positive/no/negative regulatory effects on gene $j$. $X_i(t)$ is the expression level of gene $i$ at time $t$, and $W_j(t)$ is a standard Brownian motion. {For different genes, the corresponding Brownian motions are independent.} {The same as in previous papers \cite{pinna2010knockouts,papili2018sincerities,wang2024gene}, the} parameter values are $V=30$, $\beta=1$, $\theta=0.2$, and $\sigma=0.1$. Our goal is to infer $A_\Z$ from $X_j(t)$. 

Most deep learning-based GRN inference methods are trained on experimental data sets, whose amount is very limited. Therefore, such methods need to carefully choose the neural network structure and the training procedure under this limitation. In this paper, we consider a scenario in which the training data set is sufficiently large, and study how deep learning can achieve the best performance without data limitation. 

\section{Results}

\subsection{TRENDY method}

WENDY method \cite{wang2024gene} uses single-cell gene expression data measured at two time points, each of size $m\times n$ (mRNA counts of $n$ genes for $m$ cells). Here we do not know how cells at different time points correspond to each other, since the measurement of single-cell level gene expression needs to kill the cells \cite{wang2022inference}. For data at two time points, the cell numbers $m$ can be different.

For gene expression data at time points $0$ and $t$, WENDY applies graphical lasso \cite{friedman2008sparse} to calculate the covariance matrices for different genes, $K_0$ and $K_t$, each of size $n\times n$. We can construct a general SDE system like Eq.~\ref{eqnl} to model the dynamics of gene regulation. After linearization of this system, one can obtain an approximated relation for $K_0,K_t$, and the ground truth GRN $A_\Z$, also of size $n\times n$:
\begin{equation}
    K_t= (I+tA_\Z^\T)K_0(I+tA_\Z)+D+E.
    \label{k0kt}
\end{equation}
Here $I$ is an $n\times n$ identity matrix, $D$ is an unknown $n\times n$ diagonal matrix, and $E$ is the unknown error introduced by linearization. Then WENDY solves the GRN matrix $A$ as a non-convex optimization problem:
\begin{equation}
    \arg\min_A f(A) := \frac{1}{2}\sum_{i\ne j} \{[K_t-(I+tA^\T)K_0(I+tA)]_{i,j}\}^2.
    \label{opt}
\end{equation}
Due to the unknown diagonal matrix $D$, this summation does not count elements with $i=j$. The result of this optimization problem is denoted as 
\[A_0 = \text{WENDY}(K_0,K_t).\]
{ See Section~S1 for details of WENDY.}

If WENDY works perfectly, then 
\[K_t= (I+tA_0^\T)K_0(I+tA_0)+D,\]
meaning that $(I+tA_\Z^\T)K_0(I+tA_\Z)$ and $(I+tA_0^\T)K_0(I+tA_0)$ differ by $E$, the error introduced by the linearization, and $A_0$ may not match $A_\Z$ accurately. If we want to derive a more accurate and complicated equation of $K_0$, $K_t$, and $A_\Z$ than Eq.~\ref{k0kt}, this equation might be numerically unstable to solve. Therefore, we need to find another way to improve WENDY. 

Define 
\begin{equation}
    K_t^* = (I+tA_\Z^\T)K_0(I+tA_\Z).
    \label{ktstar}
\end{equation}
If we replace $K_t$ by $K_t^*$ in Eq.~\ref{opt} and solve $A$, then this inferred $A$ should be very close to $A_\Z$. However, in practice, we only have $K_0$ and $K_t$, but not $K_t^*$, since $A_\Z$ is unknown. 

The core idea is to train a transformer model $\mathrm{TE}(k=1)$ with input $K_t$, so that the output $K_t'$ is close to $K_t^*$. Then we can calculate
\[A_1 = \text{WENDY}(K_0,K_t'),\]
which is closer to $A_\Z$ than $A_0$. We can further improve $A_1$ using another transformer model $\mathrm{TE}(k=3)$, along with the information in $K_0,K_t$, so that the output $A_2$ is even closer to $A_\Z$. These two transformer models will be discussed in the next subsection.

{ The actual $K_t$ is a complicated nonlinear function of $A_\Z$ and $K_0$. Thus $K_t^*$, also a function of $A_\Z$ and $K_0$, can be roughly learned by the $\mathrm{TE}(k=1)$ model as a function of $K_t$. Given accurate $K_0$ and $K_t^*$, we still cannot uniquely determine $A_\Z$, and the numerical solver for Eq.~\ref{opt} determines which $A_1$ is chosen. Therefore, the $\mathrm{TE}(k=3)$ model can partially learn the nonlinear behavior of the solver and output a better $A_2$ from $A_1$.}

{ Fig.~\ref{flow}, Algorithm~\ref{alg1}, and Algorithm~\ref{alg2} describe the training and testing workflow for TRENDY. }

{ \begin{algorithm}[!htbp]
	\caption{Training workflow of TRENDY method.}
	\label{alg1}
	\ \\
	\begin{enumerate}
		{	\item \textbf{Repeat} generating random $A_\Z$ and corresponding gene expression data
			
			\item \textbf{Calculate} covariance matrices $K_0$ and $K_t$, and then calculate $K_t^*$ from $K_0,K_t,A_\Z$

                \item \textbf{Train} $\mathrm{TE}(k=1)$ model with input $K_t$ and target $K_t^*$

                \textbf{Call} trained $\mathrm{TE}(k=1)$ model to calculate $K_t'$ from $K_t$
		
			\item \textbf{Call} WENDY to calculate $A_1$ from $K_0,K_t'$ 

			\item \textbf{Train} $\mathrm{TE}(k=3)$ model with input $A_1,K_0,K_t$ and target $A_\Z$
			
		}
	\end{enumerate}
\end{algorithm}}

{ \begin{algorithm}[!htbp]
	\caption{Testing workflow of TRENDY method.}
	\label{alg2}
	\ \\
	\begin{enumerate}
		{	\item \textbf{Input}: gene expression data at two time points
			
			\item \textbf{Calculate} covariance matrices $K_0$ and $K_t$

                \item \textbf{Call} trained $\mathrm{TE}(k=1)$ model to calculate $K_t'$ from $K_t$ 
		
			\item \textbf{Call} WENDY to calculate $A_1$ from $K_0,K_t'$ 

			\item \textbf{Call} trained $\mathrm{TE}(k=3)$ model to calculate $A_2$ from $A_1,K_0,K_t$

                \item \textbf{Output}: inferred GRN $A_2$
			
		}
	\end{enumerate}
\end{algorithm}}

\begin{figure}[htbp]
    \centering
    \resizebox{4.0 in}{!}{%
    \begin{tikzpicture}

\draw[line width=1.5pt,->] (6.25,10.75) -- (6.25,12.5);
\draw[line width=1.5pt,->] (11.25,6.75) -- (11.25,5.1);
\draw[line width=1.5pt,->] (1.25,6.75) -- (1.25,5.1);
\draw[line width=1.5pt,->,dashed] (11.25,12.5) -- (11.25,11.1);
\draw[line width=1.5pt,->,dashed] (1.25,12.5) -- (1.25,11.1);
\draw [line width=1pt, double distance=3pt,
             arrows = {-Latex[length=0pt 3 0]}]        (7.75,14)   -- (9.75,14);
\draw [line width=1pt, double distance=3pt,
             arrows = {-Latex[length=0pt 3 0]}]        (4.75,14)   -- (2.75,14);
\draw [line width=1pt, double distance=3pt,
             arrows = {-Latex[length=0pt 3 0]}]        (9.75,3.75)   -- (7.75,3.75);
\draw [line width=1pt, double distance=3pt,
             arrows = {-Latex[length=0pt 3 0]}]        (1.25,2.4)   -- (1.25, 1.0);
\draw [line width=1pt, double distance=3pt,
             arrows = {-Latex[length=0pt 3 0]}]        (7.75,-1.25)   -- (9.75,-1.25);
\draw[line width=1.5pt,->] (4.75,3.75) -- (2.75,3.75);
\draw[line width=1.5pt,->] (2.75,-1.25) -- (4.75,-1.25);
\draw[line width=1.5pt,->] (2.55,7.35) -- (5.25,-0.25);
\draw[line width=1.5pt,->] (9.95,7.35) -- (7.25,-0.25);

\draw[fill=cyan!30] (6.25, 14) circle [radius=1.25];
\node at (6.25, 14) {{\begin{tabular}{c} \Large data \\ \Large generator \end{tabular}}};
\node at (1.25, 15.5) {{\begin{tabular}{c} \Large expression \\ \Large data \end{tabular}}};
\node at (11.25, 15.5) {{\begin{tabular}{c} \Large expression \\ \Large data \end{tabular}}};
\node at (1.25, 12.75) {\Large at time $0$};
\node at (11.25, 12.75) {\Large at time $t$};
\node at (1.25, 7.1) {\Large $K_0$};
\node at (11.25, 7.1) {\Large $K_t$};
\node at (6.25, 10.35) {\Large true GRN};
\node at (6.25, 7.1) {\Large $A_{\mathrm{true}}$};
\node at (6.25, 5.5) {{\begin{tabular}{c} \Large revised \\ \Large covariance \end{tabular}}};
\node at (1.25, 0.5) {{\begin{tabular}{c} \Large inferred \\ \Large GRN \end{tabular}}};
\node at (11.25, 0.5) {{\begin{tabular}{c} \Large inferred \\ \Large GRN \end{tabular}}};
\node at (1.25, -2.9) {\Large $A_1$};
\node at (11.25, -2.9) {\Large $A_2$};
\node at (8.8, -0.8) {\Large \textcolor{blue}{$(5)$}};
\node at (3.8, -0.8) {\Large \textcolor{blue}{$(5)$}};
\node at (7.8, 0.1) {\Large \textcolor{blue}{$(5)$}};
\node at (4.7, 0.1) {\Large \textcolor{blue}{$(5)$}};
\node at (9.2, 3.2) {\Large \textcolor[rgb]{0, 0.7, 0.3}{$(3)$}};
\node at (11.8, 6) {\Large \textcolor[rgb]{0, 0.7, 0.3}{$(3)$}};
\node at (1.8, 2) {\Large \textcolor{cyan}{$(4)$}};
\node at (1.8, 6) {\Large \textcolor{cyan}{$(4)$}};
\node at (3.1, 3.2) {\Large \textcolor{cyan}{$(4)$}};
\node at (1.7, 11.8) {\Large \textcolor{orange}{$(2)$}};
\node at (10.8, 11.8) {\Large \textcolor{orange}{$(2)$}};
\node at (5.8, 11.8) {\Large \textcolor{red}{$(1)$}};
\node at (4, 13.5) {\Large \textcolor{red}{$(1)$}};
\node at (8.5, 13.5) {\Large \textcolor{red}{$(1)$}};
\node at (6.25, 2.1) {\Large $K_t'$};
\node at (1.25, 10.5) {{\begin{tabular}{c} \Large covariance \\ \Large matrix \end{tabular}}};
\node at (11.25, 10.5) {{\begin{tabular}{c} \Large covariance \\ \Large matrix \end{tabular}}};
\draw[fill=yellow!41!green!21] (11.25,3.75) circle [radius=1.25];
\node at (11.25,3.5) {{\begin{tabular}{c} \Large $\mathrm{TE}(k=1)$ \\ \Large model  \end{tabular}}};
\draw[fill=orange!21] (6.25,-1.25) circle [radius=1.25];
\node at (6.25,-1.5) {{\begin{tabular}{c} \Large $\mathrm{TE}(k=3)$ \\ \Large model  \end{tabular}}};
\draw[fill=orange!21] (1.25,3.75) circle [radius=1.25];
\node at (1.25,3.75) {{\begin{tabular}{c} \Large WENDY \\ \Large method \end{tabular}}};

\filldraw[fill=blue!30!green!81] (0.0,10.0) rectangle (0.5,9.5);
\filldraw[fill=blue!30!green!14] (0.5,10.0) rectangle (1.0,9.5);
\filldraw[fill=blue!30!green!3] (1.0,10.0) rectangle (1.5,9.5);
\filldraw[fill=blue!30!green!94] (1.5,10.0) rectangle (2.0,9.5);
\filldraw[fill=blue!30!green!35] (2.0,10.0) rectangle (2.5,9.5);
\filldraw[fill=blue!30!green!14] (0.0,9.5) rectangle (0.5,9.0);
\filldraw[fill=blue!30!green!28] (0.5,9.5) rectangle (1.0,9.0);
\filldraw[fill=blue!30!green!17] (1.0,9.5) rectangle (1.5,9.0);
\filldraw[fill=blue!30!green!94] (1.5,9.5) rectangle (2.0,9.0);
\filldraw[fill=blue!30!green!13] (2.0,9.5) rectangle (2.5,9.0);
\filldraw[fill=blue!30!green!3] (0.0,9.0) rectangle (0.5,8.5);
\filldraw[fill=blue!30!green!17] (0.5,9.0) rectangle (1.0,8.5);
\filldraw[fill=blue!30!green!69] (1.0,9.0) rectangle (1.5,8.5);
\filldraw[fill=blue!30!green!11] (1.5,9.0) rectangle (2.0,8.5);
\filldraw[fill=blue!30!green!75] (2.0,9.0) rectangle (2.5,8.5);
\filldraw[fill=blue!30!green!94] (0.0,8.5) rectangle (0.5,8.0);
\filldraw[fill=blue!30!green!94] (0.5,8.5) rectangle (1.0,8.0);
\filldraw[fill=blue!30!green!11] (1.0,8.5) rectangle (1.5,8.0);
\filldraw[fill=blue!30!green!11] (1.5,8.5) rectangle (2.0,8.0);
\filldraw[fill=blue!30!green!27] (2.0,8.5) rectangle (2.5,8.0);
\filldraw[fill=blue!30!green!35] (0.0,8.0) rectangle (0.5,7.5);
\filldraw[fill=blue!30!green!13] (0.5,8.0) rectangle (1.0,7.5);
\filldraw[fill=blue!30!green!75] (1.0,8.0) rectangle (1.5,7.5);
\filldraw[fill=blue!30!green!27] (1.5,8.0) rectangle (2.0,7.5);
\filldraw[fill=blue!30!green!71] (2.0,8.0) rectangle (2.5,7.5);
\filldraw[fill=blue!30!green!25] (10.0,10.0) rectangle (10.5,9.5);
\filldraw[fill=blue!30!green!91] (10.5,10.0) rectangle (11.0,9.5);
\filldraw[fill=blue!30!green!83] (11.0,10.0) rectangle (11.5,9.5);
\filldraw[fill=blue!30!green!89] (11.5,10.0) rectangle (12.0,9.5);
\filldraw[fill=blue!30!green!69] (12.0,10.0) rectangle (12.5,9.5);
\filldraw[fill=blue!30!green!91] (10.0,9.5) rectangle (10.5,9.0);
\filldraw[fill=blue!30!green!28] (10.5,9.5) rectangle (11.0,9.0);
\filldraw[fill=blue!30!green!57] (11.0,9.5) rectangle (11.5,9.0);
\filldraw[fill=blue!30!green!75] (11.5,9.5) rectangle (12.0,9.0);
\filldraw[fill=blue!30!green!35] (12.0,9.5) rectangle (12.5,9.0);
\filldraw[fill=blue!30!green!83] (10.0,9.0) rectangle (10.5,8.5);
\filldraw[fill=blue!30!green!57] (10.5,9.0) rectangle (11.0,8.5);
\filldraw[fill=blue!30!green!20] (11.0,9.0) rectangle (11.5,8.5);
\filldraw[fill=blue!30!green!89] (11.5,9.0) rectangle (12.0,8.5);
\filldraw[fill=blue!30!green!54] (12.0,9.0) rectangle (12.5,8.5);
\filldraw[fill=blue!30!green!89] (10.0,8.5) rectangle (10.5,8.0);
\filldraw[fill=blue!30!green!75] (10.5,8.5) rectangle (11.0,8.0);
\filldraw[fill=blue!30!green!89] (11.0,8.5) rectangle (11.5,8.0);
\filldraw[fill=blue!30!green!27] (11.5,8.5) rectangle (12.0,8.0);
\filldraw[fill=blue!30!green!97] (12.0,8.5) rectangle (12.5,8.0);
\filldraw[fill=blue!30!green!69] (10.0,8.0) rectangle (10.5,7.5);
\filldraw[fill=blue!30!green!35] (10.5,8.0) rectangle (11.0,7.5);
\filldraw[fill=blue!30!green!54] (11.0,8.0) rectangle (11.5,7.5);
\filldraw[fill=blue!30!green!97] (11.5,8.0) rectangle (12.0,7.5);
\filldraw[fill=blue!30!green!12] (12.0,8.0) rectangle (12.5,7.5);
\filldraw[fill=blue!30!green!27] (5.0,5.0) rectangle (5.5,4.5);
\filldraw[fill=blue!30!green!93] (5.5,5.0) rectangle (6.0,4.5);
\filldraw[fill=blue!30!green!100] (6.0,5.0) rectangle (6.5,4.5);
\filldraw[fill=blue!30!green!85] (6.5,5.0) rectangle (7.0,4.5);
\filldraw[fill=blue!30!green!51] (7.0,5.0) rectangle (7.5,4.5);
\filldraw[fill=blue!30!green!93] (5.0,4.5) rectangle (5.5,4.0);
\filldraw[fill=blue!30!green!42] (5.5,4.5) rectangle (6.0,4.0);
\filldraw[fill=blue!30!green!44] (6.0,4.5) rectangle (6.5,4.0);
\filldraw[fill=blue!30!green!79] (6.5,4.5) rectangle (7.0,4.0);
\filldraw[fill=blue!30!green!20] (7.0,4.5) rectangle (7.5,4.0);
\filldraw[fill=blue!30!green!100] (5.0,4.0) rectangle (5.5,3.5);
\filldraw[fill=blue!30!green!44] (5.5,4.0) rectangle (6.0,3.5);
\filldraw[fill=blue!30!green!40] (6.0,4.0) rectangle (6.5,3.5);
\filldraw[fill=blue!30!green!100] (6.5,4.0) rectangle (7.0,3.5);
\filldraw[fill=blue!30!green!57] (7.0,4.0) rectangle (7.5,3.5);
\filldraw[fill=blue!30!green!85] (5.0,3.5) rectangle (5.5,3.0);
\filldraw[fill=blue!30!green!79] (5.5,3.5) rectangle (6.0,3.0);
\filldraw[fill=blue!30!green!100] (6.0,3.5) rectangle (6.5,3.0);
\filldraw[fill=blue!30!green!9] (6.5,3.5) rectangle (7.0,3.0);
\filldraw[fill=blue!30!green!91] (7.0,3.5) rectangle (7.5,3.0);
\filldraw[fill=blue!30!green!51] (5.0,3.0) rectangle (5.5,2.5);
\filldraw[fill=blue!30!green!20] (5.5,3.0) rectangle (6.0,2.5);
\filldraw[fill=blue!30!green!57] (6.0,3.0) rectangle (6.5,2.5);
\filldraw[fill=blue!30!green!91] (6.5,3.0) rectangle (7.0,2.5);
\filldraw[fill=blue!30!green!16] (7.0,3.0) rectangle (7.5,2.5);
\filldraw[fill=red!35] (5.0,10.0) rectangle (5.5,9.5);
\filldraw[fill=red!58] (5.5,10.0) rectangle (6.0,9.5);
\filldraw[fill=red!81] (6.0,10.0) rectangle (6.5,9.5);
\filldraw[fill=red!46] (6.5,10.0) rectangle (7.0,9.5);
\filldraw[fill=red!20] (7.0,10.0) rectangle (7.5,9.5);
\filldraw[fill=red!47] (5.0,9.5) rectangle (5.5,9.0);
\filldraw[fill=red!45] (5.5,9.5) rectangle (6.0,9.0);
\filldraw[fill=red!26] (6.0,9.5) rectangle (6.5,9.0);
\filldraw[fill=red!85] (6.5,9.5) rectangle (7.0,9.0);
\filldraw[fill=red!34] (7.0,9.5) rectangle (7.5,9.0);
\filldraw[fill=red!89] (5.0,9.0) rectangle (5.5,8.5);
\filldraw[fill=red!87] (5.5,9.0) rectangle (6.0,8.5);
\filldraw[fill=red!82] (6.0,9.0) rectangle (6.5,8.5);
\filldraw[fill=red!9] (6.5,9.0) rectangle (7.0,8.5);
\filldraw[fill=red!77] (7.0,9.0) rectangle (7.5,8.5);
\filldraw[fill=red!81] (5.0,8.5) rectangle (5.5,8.0);
\filldraw[fill=red!21] (5.5,8.5) rectangle (6.0,8.0);
\filldraw[fill=red!68] (6.0,8.5) rectangle (6.5,8.0);
\filldraw[fill=red!93] (6.5,8.5) rectangle (7.0,8.0);
\filldraw[fill=red!31] (7.0,8.5) rectangle (7.5,8.0);
\filldraw[fill=red!20] (5.0,8.0) rectangle (5.5,7.5);
\filldraw[fill=red!59] (5.5,8.0) rectangle (6.0,7.5);
\filldraw[fill=red!48] (6.0,8.0) rectangle (6.5,7.5);
\filldraw[fill=red!34] (6.5,8.0) rectangle (7.0,7.5);
\filldraw[fill=red!81] (7.0,8.0) rectangle (7.5,7.5);
\filldraw[fill=red!49] (0.0,0.0) rectangle (0.5,-0.5);
\filldraw[fill=red!63] (0.5,0.0) rectangle (1.0,-0.5);
\filldraw[fill=red!65] (1.0,0.0) rectangle (1.5,-0.5);
\filldraw[fill=red!59] (1.5,0.0) rectangle (2.0,-0.5);
\filldraw[fill=red!10] (2.0,0.0) rectangle (2.5,-0.5);
\filldraw[fill=red!70] (0.0,-0.5) rectangle (0.5,-1.0);
\filldraw[fill=red!64] (0.5,-0.5) rectangle (1.0,-1.0);
\filldraw[fill=red!45] (1.0,-0.5) rectangle (1.5,-1.0);
\filldraw[fill=red!58] (1.5,-0.5) rectangle (2.0,-1.0);
\filldraw[fill=red!18] (2.0,-0.5) rectangle (2.5,-1.0);
\filldraw[fill=red!100] (0.0,-1.0) rectangle (0.5,-1.5);
\filldraw[fill=red!59] (0.5,-1.0) rectangle (1.0,-1.5);
\filldraw[fill=red!100] (1.0,-1.0) rectangle (1.5,-1.5);
\filldraw[fill=red!0] (1.5,-1.0) rectangle (2.0,-1.5);
\filldraw[fill=red!72] (2.0,-1.0) rectangle (2.5,-1.5);
\filldraw[fill=red!68] (0.0,-1.5) rectangle (0.5,-2.0);
\filldraw[fill=red!0] (0.5,-1.5) rectangle (1.0,-2.0);
\filldraw[fill=red!51] (1.0,-1.5) rectangle (1.5,-2.0);
\filldraw[fill=red!100] (1.5,-1.5) rectangle (2.0,-2.0);
\filldraw[fill=red!61] (2.0,-1.5) rectangle (2.5,-2.0);
\filldraw[fill=red!26] (0.0,-2.0) rectangle (0.5,-2.5);
\filldraw[fill=red!85] (0.5,-2.0) rectangle (1.0,-2.5);
\filldraw[fill=red!63] (1.0,-2.0) rectangle (1.5,-2.5);
\filldraw[fill=red!24] (1.5,-2.0) rectangle (2.0,-2.5);
\filldraw[fill=red!64] (2.0,-2.0) rectangle (2.5,-2.5);
\filldraw[fill=red!45] (10.0,0.0) rectangle (10.5,-0.5);
\filldraw[fill=red!63] (10.5,0.0) rectangle (11.0,-0.5);
\filldraw[fill=red!83] (11.0,0.0) rectangle (11.5,-0.5);
\filldraw[fill=red!56] (11.5,0.0) rectangle (12.0,-0.5);
\filldraw[fill=red!24] (12.0,0.0) rectangle (12.5,-0.5);
\filldraw[fill=red!41] (10.0,-0.5) rectangle (10.5,-1.0);
\filldraw[fill=red!43] (10.5,-0.5) rectangle (11.0,-1.0);
\filldraw[fill=red!20] (11.0,-0.5) rectangle (11.5,-1.0);
\filldraw[fill=red!82] (11.5,-0.5) rectangle (12.0,-1.0);
\filldraw[fill=red!41] (12.0,-0.5) rectangle (12.5,-1.0);
\filldraw[fill=red!96] (10.0,-1.0) rectangle (10.5,-1.5);
\filldraw[fill=red!85] (10.5,-1.0) rectangle (11.0,-1.5);
\filldraw[fill=red!90] (11.0,-1.0) rectangle (11.5,-1.5);
\filldraw[fill=red!12] (11.5,-1.0) rectangle (12.0,-1.5);
\filldraw[fill=red!85] (12.0,-1.0) rectangle (12.5,-1.5);
\filldraw[fill=red!83] (10.0,-1.5) rectangle (10.5,-2.0);
\filldraw[fill=red!22] (10.5,-1.5) rectangle (11.0,-2.0);
\filldraw[fill=red!65] (11.0,-1.5) rectangle (11.5,-2.0);
\filldraw[fill=red!87] (11.5,-1.5) rectangle (12.0,-2.0);
\filldraw[fill=red!37] (12.0,-1.5) rectangle (12.5,-2.0);
\filldraw[fill=red!25] (10.0,-2.0) rectangle (10.5,-2.5);
\filldraw[fill=red!51] (10.5,-2.0) rectangle (11.0,-2.5);
\filldraw[fill=red!39] (11.0,-2.0) rectangle (11.5,-2.5);
\filldraw[fill=red!27] (11.5,-2.0) rectangle (12.0,-2.5);
\filldraw[fill=red!75] (12.0,-2.0) rectangle (12.5,-2.5);
\filldraw[fill=blue!80] (0.0,15.0) rectangle (0.5,14.5);
\filldraw[fill=blue!20] (0.5,15.0) rectangle (1.0,14.5);
\filldraw[fill=blue!87] (1.0,15.0) rectangle (1.5,14.5);
\filldraw[fill=blue!54] (1.5,15.0) rectangle (2.0,14.5);
\filldraw[fill=blue!76] (2.0,15.0) rectangle (2.5,14.5);
\filldraw[fill=blue!8] (0.0,14.5) rectangle (0.5,14.0);
\filldraw[fill=blue!49] (0.5,14.5) rectangle (1.0,14.0);
\filldraw[fill=blue!48] (1.0,14.5) rectangle (1.5,14.0);
\filldraw[fill=blue!76] (1.5,14.5) rectangle (2.0,14.0);
\filldraw[fill=blue!59] (2.0,14.5) rectangle (2.5,14.0);
\filldraw[fill=blue!67] (0.0,14.0) rectangle (0.5,13.5);
\filldraw[fill=blue!32] (0.5,14.0) rectangle (1.0,13.5);
\filldraw[fill=blue!70] (1.0,14.0) rectangle (1.5,13.5);
\filldraw[fill=blue!1] (1.5,14.0) rectangle (2.0,13.5);
\filldraw[fill=blue!87] (2.0,14.0) rectangle (2.5,13.5);
\filldraw[fill=blue!92] (0.0,13.5) rectangle (0.5,13.0);
\filldraw[fill=blue!14] (0.5,13.5) rectangle (1.0,13.0);
\filldraw[fill=blue!87] (1.0,13.5) rectangle (1.5,13.0);
\filldraw[fill=blue!68] (1.5,13.5) rectangle (2.0,13.0);
\filldraw[fill=blue!96] (2.0,13.5) rectangle (2.5,13.0);
\filldraw[fill=blue!34] (10.0,15.0) rectangle (10.5,14.5);
\filldraw[fill=blue!98] (10.5,15.0) rectangle (11.0,14.5);
\filldraw[fill=blue!82] (11.0,15.0) rectangle (11.5,14.5);
\filldraw[fill=blue!43] (11.5,15.0) rectangle (12.0,14.5);
\filldraw[fill=blue!14] (12.0,15.0) rectangle (12.5,14.5);
\filldraw[fill=blue!37] (10.0,14.5) rectangle (10.5,14.0);
\filldraw[fill=blue!55] (10.5,14.5) rectangle (11.0,14.0);
\filldraw[fill=blue!20] (11.0,14.5) rectangle (11.5,14.0);
\filldraw[fill=blue!58] (11.5,14.5) rectangle (12.0,14.0);
\filldraw[fill=blue!0] (12.0,14.5) rectangle (12.5,14.0);
\filldraw[fill=blue!92] (10.0,14.0) rectangle (10.5,13.5);
\filldraw[fill=blue!92] (10.5,14.0) rectangle (11.0,13.5);
\filldraw[fill=blue!33] (11.0,14.0) rectangle (11.5,13.5);
\filldraw[fill=blue!64] (11.5,14.0) rectangle (12.0,13.5);
\filldraw[fill=blue!97] (12.0,14.0) rectangle (12.5,13.5);
\filldraw[fill=blue!22] (10.0,13.5) rectangle (10.5,13.0);
\filldraw[fill=blue!64] (10.5,13.5) rectangle (11.0,13.0);
\filldraw[fill=blue!13] (11.0,13.5) rectangle (11.5,13.0);
\filldraw[fill=blue!80] (11.5,13.5) rectangle (12.0,13.0);
\filldraw[fill=blue!38] (12.0,13.5) rectangle (12.5,13.0);
\draw[dashed] (3,6.8) rectangle (9.5,15.5);

    \end{tikzpicture}
}
    \caption{Workflow of training and testing the TRENDY method. Rectangles with grids are numerical matrices, where the color scale represents the numerical value. Circles are mechanisms that take in matrices and produce matrices. Single and double arrows represent inputs and outputs of each mechanism, and dashed arrows represent direct calculations. To apply TRENDY after training, the step in the dashed rectangle is omitted.}
    \label{flow}
\end{figure}

\subsection{Transformer model $\mathrm{TE}(k)$}
We build a deep learning framework $\mathrm{TE}(k)$ that is based on transformer encoder layers, where $k$ is the major hyperparameter that describes the number of input matrices. Besides, there are three hyperparameters that can be tuned: the model dimension $d$; the number of encoder layers $l$; the number of attention heads $h$ of the encoder layer. We use this $\mathrm{TE}(k)$ model with different hyperparameters in TRENDY and other transformer-enhanced methods. See Algorithm~\ref{alg3} for the general structure of $\mathrm{TE}(k)$, along with the shape of data after each layer. { Besides the standard transformer structure, the segment embedding layer tells the model to treat $k>1$ input matrices separately, and the 2-D positional encoding layer tells the model to remember that the input is originally two-dimensional.} Notice that the same transformer encoder layer can handle inputs of different lengths. This means that the gene number $n$ is not a predetermined hyperparameter, and we do not need to train a different model for each $n$. We will train the $\mathrm{TE}(k)$ model with $n=10$ genes and test on data sets with $n=10/18/20$ genes.

For the first half of the TRENDY method, the $\mathrm{TE}(k=1)$ model has $k=1$ input matrix $K_t$, $d=64$ model dimension, $l=7$ encoder layers, and $h=4$ attention heads. For the second half of the TRENDY method, the $\mathrm{TE}(k=3)$ model has $k=3$ input matrices of the same size, $A_1,K_0,K_t$, $d=64$ model dimension, $l=7$ encoder layers, and $h=8$ attention heads.

{ \begin{algorithm}[!htbp]
	\caption{Structure of the $\mathrm{TE}(k)$ model. The shape of data after each layer is in the brackets.}
	\label{alg3}
	\ \\
	\begin{enumerate}
		{	\item \textbf{Input}: $k$ matrices ($k$ groups of $n\times n$)
			
			\item Linear embedding layer with dimension $d$ ($k$ groups of $n\times n \times d$)

                \item Segment embedding layer (omitted when $k=1$) ($k$ groups of $n\times n \times d$)
		
			\item 2-D positional encoding layer ($k$ groups of $n\times n \times d$)

			\item Flattening and concatenation ($(n^2) \times (dk)$)

                \item $l$ layers of transformer encoder ($(n^2) \times (dk)$)

                \item Linear embedding layer ($n^2 \times 1$)

                \item \textbf{Output}: reshaping into a matrix ($n \times n$)
			
		}
	\end{enumerate}
\end{algorithm}}

\subsection{Other transformer-enhanced methods}
{ The second half of the TRENDY method applies a transformer model to learn the highly nonlinear relation between $A_1$ (result of WENDY) and $A_\Z$. For other GRN inference methods, we can assume that the inferred GRN and $A_\Z$ have some nonlinear and possibly random relations, since these methods only consider and infer certain forms of regulations. Then we can try to enhance the inferred results by a transformer model.}

GENIE3 \cite{huynh2010inferring} works on single-cell expression data at one time point. It uses random forest to select genes that have predictability for the target gene, thus being information-based.

SINCERITIES \cite{papili2018sincerities} works on single-cell expression data at multiple time points. It runs regression on the {Kolmogorov–Smirnov distance between cumulative distribution functions} of expression levels, also being information-based.

NonlinearODEs \cite{ma2020inference} works on bulk expression data at multiple time points. This method fits the data with a nonlinear differential equation system, which is thus mechanism-based.

For each of these three methods, we train a $\mathrm{TE}(k)$ model to enhance the inferred results, similar to the $\mathrm{TE}(k=3)$ model in the second half of TRENDY. Here each $\mathrm{TE}(k)$ model has $d=64,l=7,h=8$, and the training target is $A_\Z$.

For inferred GRN $A_{\mathrm{G}}$ by GENIE3, we train a $\mathrm{TE}(k=2)$ model with $A_{\mathrm{G}}$ and the covariance matrix $K$ as input. Then we can use this trained model to calculate a more accurate $A_\mathrm{G}'$ from $A_{\mathrm{G}}$ and $K$. This method is named tGENIE3.

For inferred GRN $A_{\mathrm{S}}$ by SINCERITIES, we train a $\mathrm{TE}(k=1)$ model with $A_{\mathrm{S}}$ as input, and name it tSINCERITIES.

For inferred GRN $A_{\mathrm{N}}$ by NonlinearODEs, we train a $\mathrm{TE}(k=1)$ model with $A_{\mathrm{N}}$ as input, and name it tNonlinearODEs.

\subsection{Methods enhanced by ND and BC}
{Network deconvolution (ND) \cite{feizi2013network} requires the input to be nonnegative and symmetric, so as its output. The idea is to assume that the input is a convolution of a simpler network, and solves this simpler network by deconvolution. BRANE Cut (BC) \cite{pirayre2015brane} performs well when we know that some genes are transcription factors and others are not. When we have no knowledge of transcription factors, BC degenerates to setting a threshold for whether one regulation exists. BC requires the input to be nonnegative and symmetric, and the output is a 0-1 matrix. We use ND and BC to enhance those four traditional methods, and name them as nWENDY, nGENIE3, nSINCERITIES, nNonlinearODEs, bWENDY, bGENIE3, bSINCERITIES, bNonlinearODEs.}

\subsection{Performance of different methods}

{ Certain autoregulations can decrease the variance \cite{ramos2019physical,giovanini2020comparative,holehouse2020stochastic}, and mRNA bursts can increase the gene expression variance \cite{pare2009visualization,bothma2014dynamic,leyes2023transcriptional}, which can be approximated by tuning the value of $\sigma$ in Eq.~\ref{eqnl}. SINC data set is generated by Eq.~\ref{eqnl} with $\sigma=0.01/0.1/1$. The synthetic DREAM4 data set \cite{marbach2012wisdom} is commonly used as a benchmark for GRN inference \cite{wang2024gene,papili2018sincerities}. THP-1 data set measures monocytic THP-1 human myeloid leukemia cells \cite{kouno2013temporal}. hESC data set measures human embryonic stem cell-derived progenitor cells\cite{chu2016single}. }

{ For these four data sets, two synthetic and two experimental, we test 16 methods: four traditional methods (WENDY, GENIE3, SINCERITIES, and NonlinearODEs), their transformer variants, ND variants, and BC variants. To compare the inferred GRN $A_{\mathrm{pred}}$ and the ground truth GRN $A_\Z$, we adopt the common practice that calculates two area under curve (AUC) scores: AUROC and AUPRC \cite{wang2024gene}. Both AUC scores are between $0$ and $1$, where $1$ means perfect matching, and $0$ means perfect mismatching. See Section~S2 for details of these two scores. }

\begin{figure}
    \centering
        \includegraphics[width=0.8\textwidth]{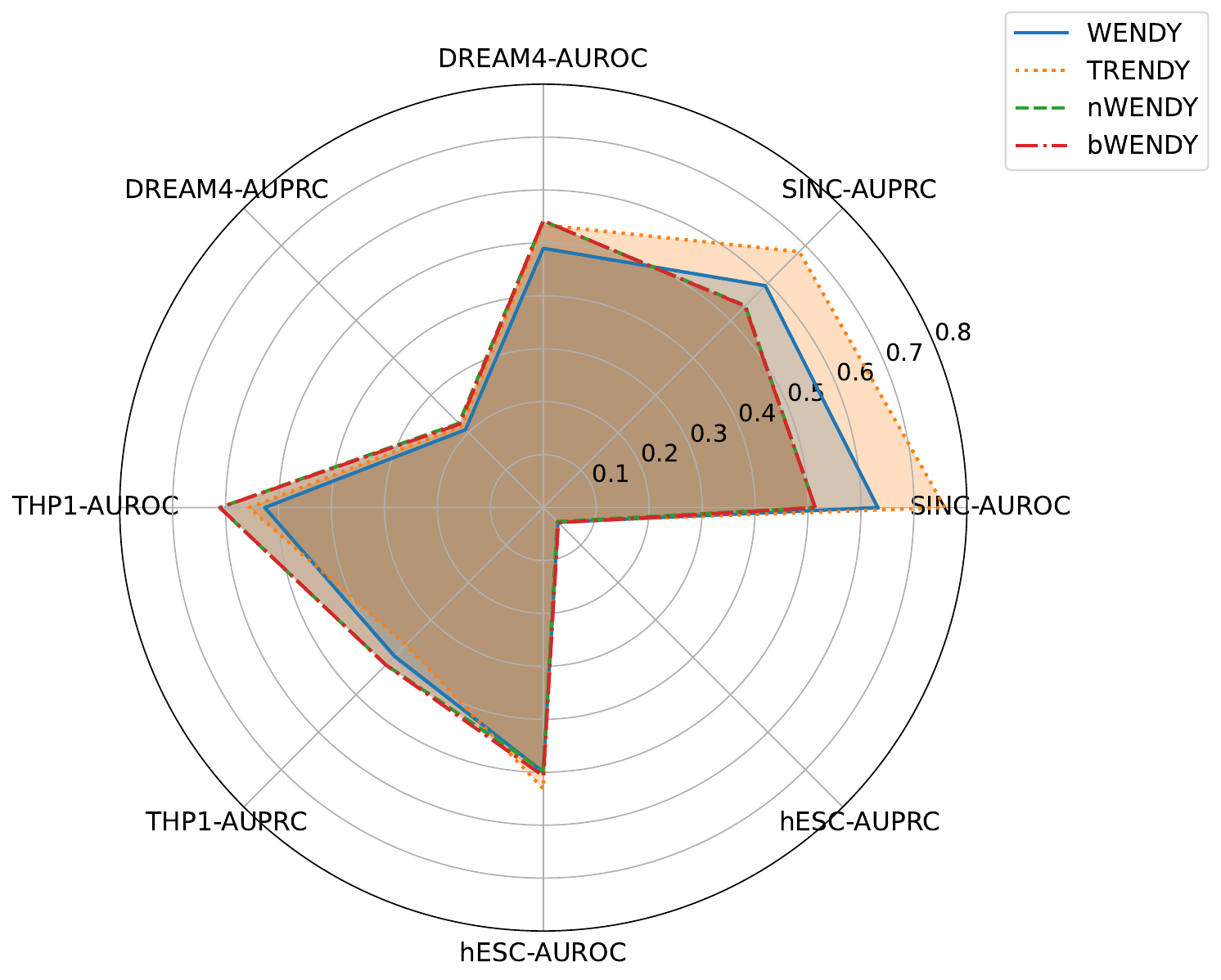}
    \caption{AUROC and AUPRC scores of WENDY and its variants on four data sets.}
    \label{wp}
\end{figure}

\begin{figure}
    \centering
        \includegraphics[width=0.8\textwidth]{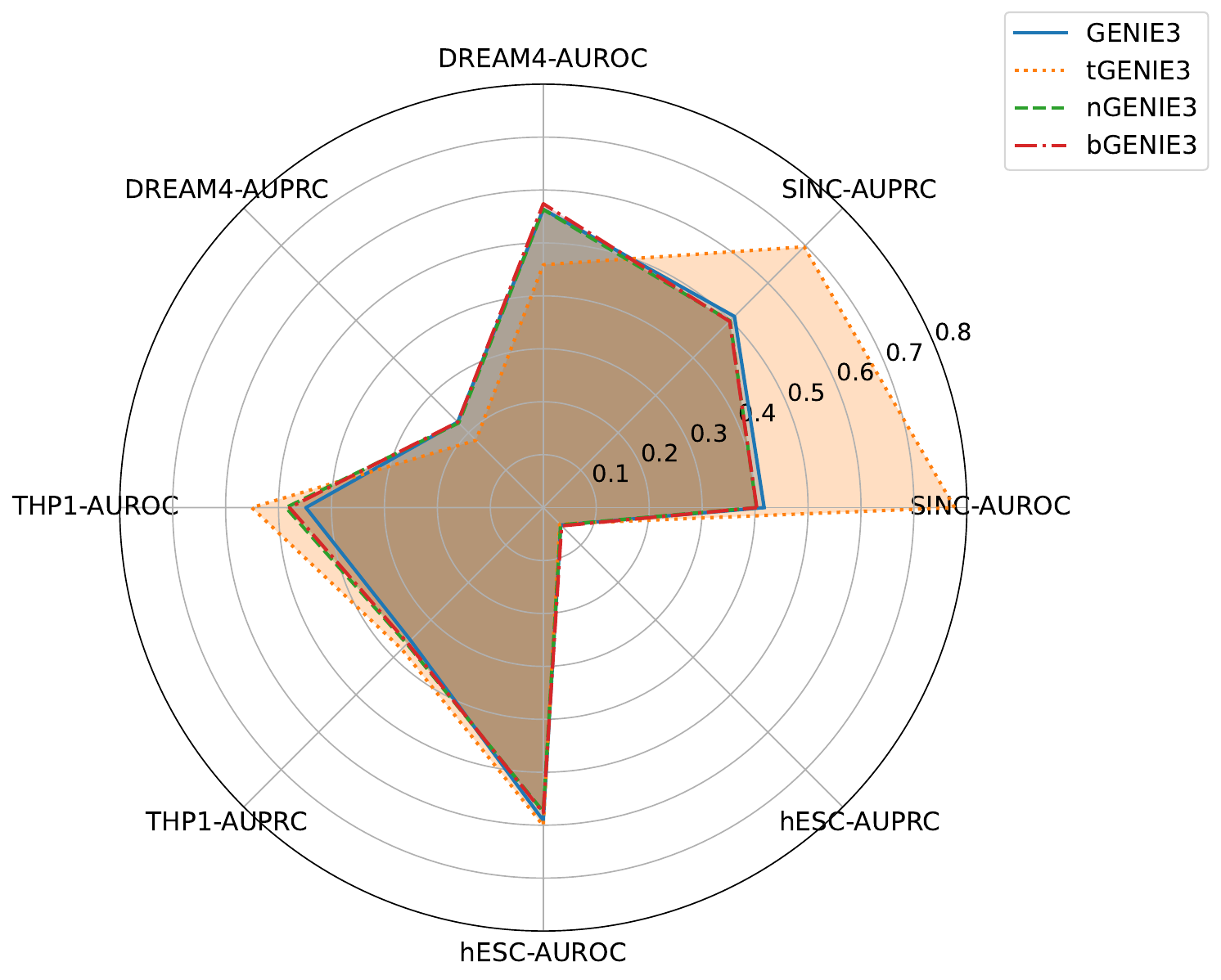}
    \caption{AUROC and AUPRC scores of GENIE3 and its variants on four data sets.}
    \label{gp}
\end{figure}

\begin{figure}
    \centering
        \includegraphics[width=0.8\textwidth]{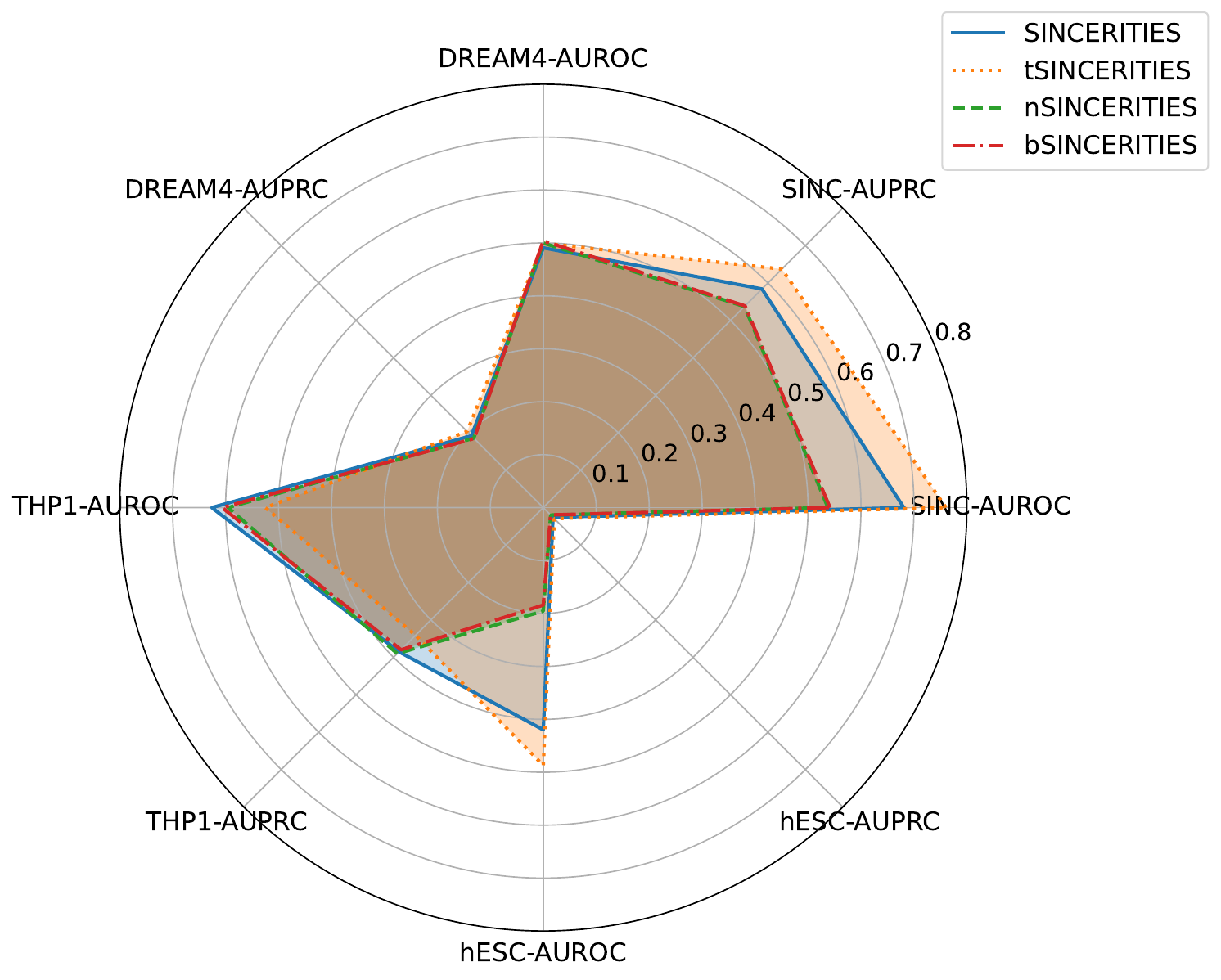}
    \caption{AUROC and AUPRC scores of SINCERITIES and its variants on four data sets.}
    \label{sp}
\end{figure}

\begin{figure}
    \centering
        \includegraphics[width=0.8\textwidth]{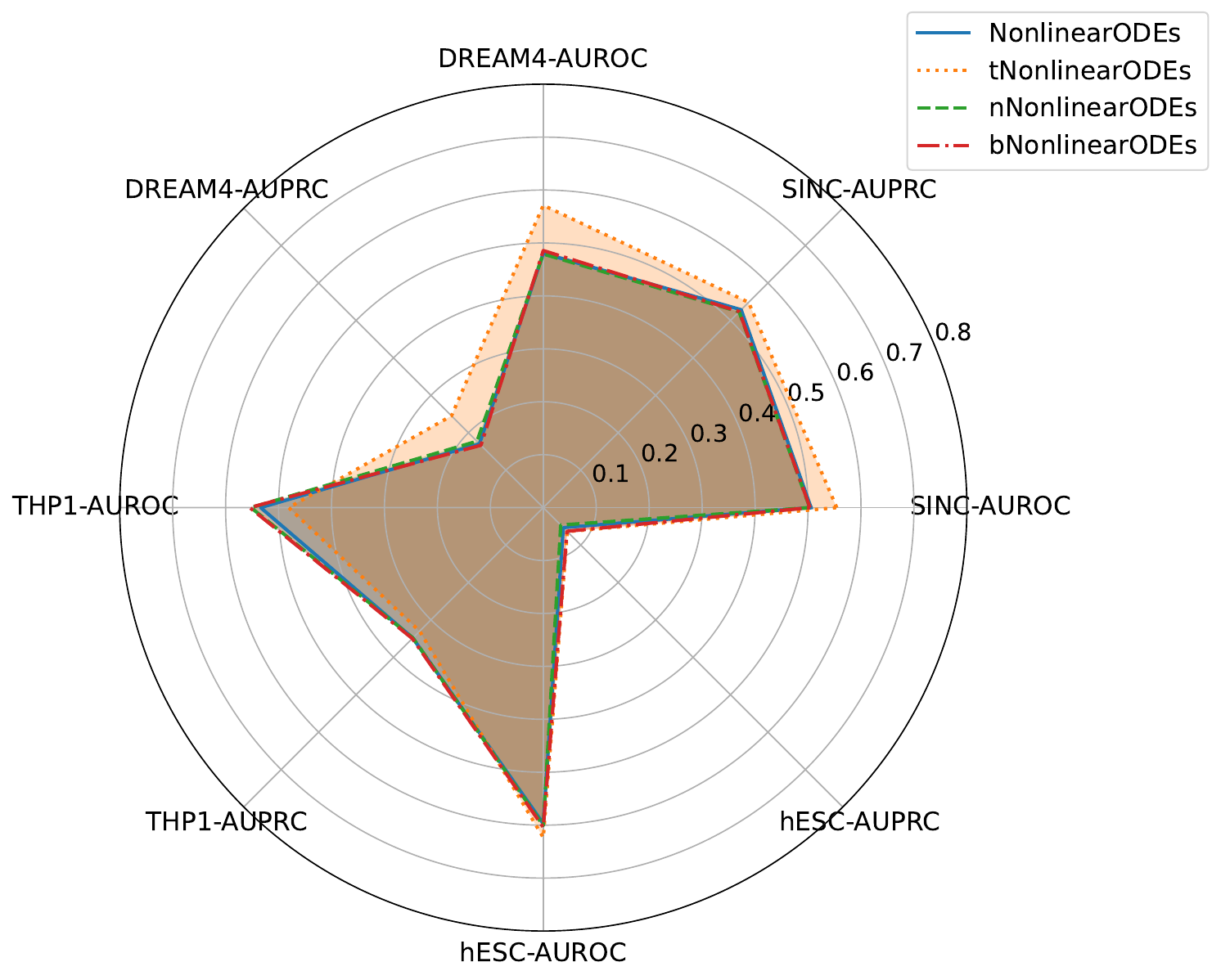}
    \caption{AUROC and AUPRC scores of NonlinearODEs and its variants on four data sets.}
    \label{np}
\end{figure}

\begin{figure}
    \centering
        \includegraphics[width=0.8\textwidth]{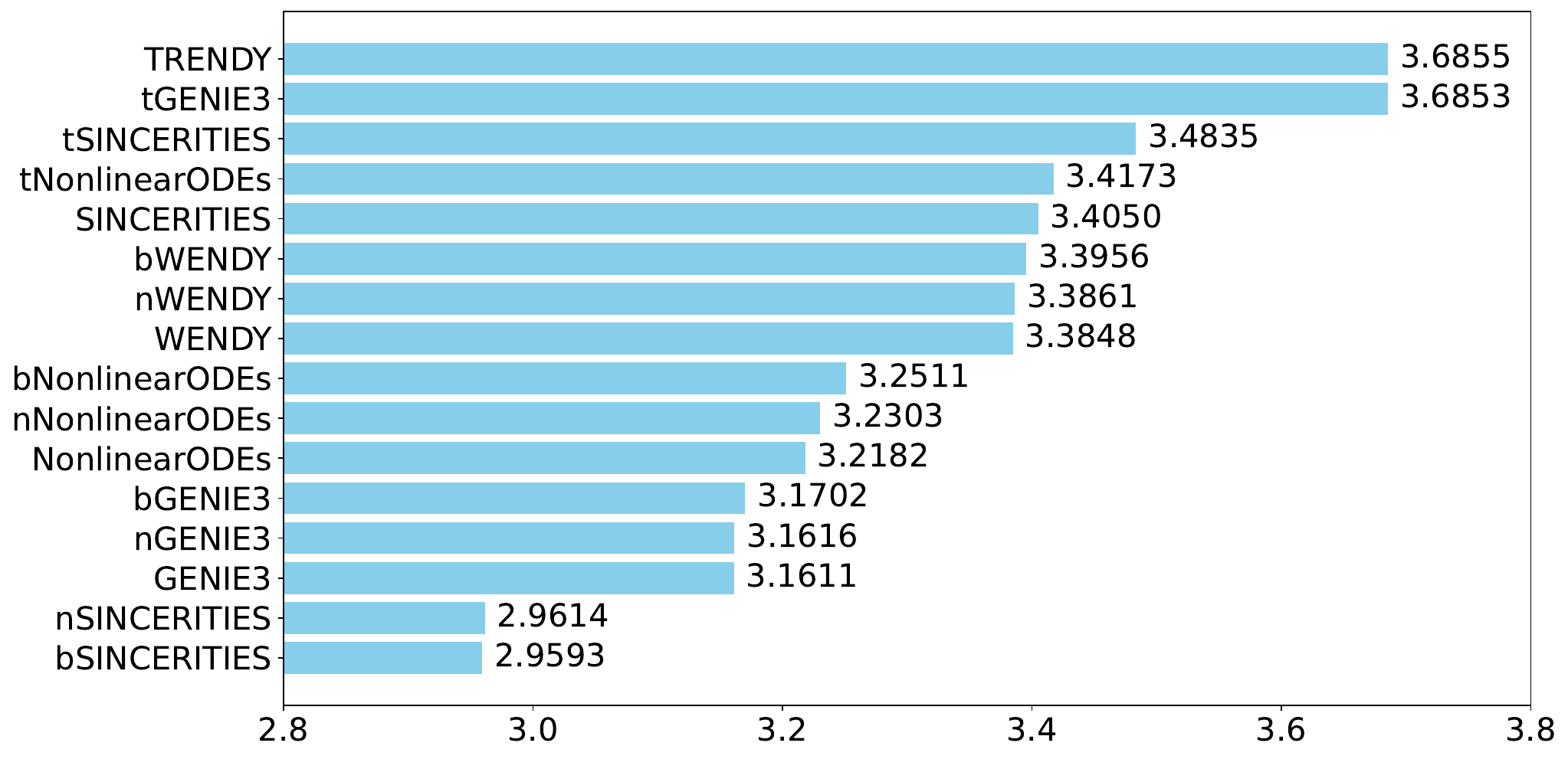}
    \caption{Total scores of all 16 methods, where the $x$-axis starts at 2.8.}
    \label{all}
\end{figure}

{ See Figs.~\ref{wp}--\ref{all} and Tables~S1,S2 for the performance of these methods.} Although the training data only has $n=10$ genes, and the THP-1 and hESC data sets have $n=20$ and $n=18$ genes, the four transformer-enhanced methods (TRENDY, tGENIE3, tSINCERITIES, tNonlinearODEs) are better than all other methods. This means that our idea of enhancing GRN inference methods with $\mathrm{TE}(k)$ models works well universally. { ND and BC generally do not enhance the original method significantly.}

The core method of this paper, TRENDY, ranks the first among all 16 methods (tGENIE3 is slightly inferior). Besides, the TRENDY method is better than the WENDY method on three data set, and has almost the same performance as WENDY on one data set (THP-1). The other three transformer-enhanced methods are worse than their non-transformer counterparts on at least one data set. This means that TRENDY's performance is robust on different data sets. In sum, TRENDY has satisfactory performance among all methods tested.

{ All methods have very small AUPRC values on hESC data set. The reason is that the ground truth GRN $A_{\mathrm{true}}$ has very few nonzero values (Fig.~S3).}

\section{Discussion}

In this paper, we present the TRENDY method, which uses the transformer model $\mathrm{TE}(k)$ to enhance the mechanism-based GRN inference method WENDY. TRENDY is tested against three other transformer-enhanced methods and their original forms, and ranks the first. This work explores the potential of deep learning methods in GRN inference when there are sufficient training data. Besides, TRENDY is developed on the biological dynamics of gene regulation, and it is more interpretable than other deep learning-based methods that treat GRN inference as a pure prediction task. { Nevertheless, the transformer encoder is directly used as an oracle machine in TRENDY. A future direction is to modify the neural network structure to better represent the biological structure \cite{liu2020fully}.}

The essential difficulty of GRN inference is the lack of data with experimentally verified GRNs. Many deep learning models need significantly more data with known GRNs, and we have to generate new data. Since we do not fully understand the relation between the dynamics of gene expression and the ground truth GRN, it is difficult to evaluate the quality of generated data along with a random GRN. We also cannot guarantee that GRN inference methods that perform well on known experimental data sets can still be advantageous if there are many more experimental data with different GRNs.

We train the models with data of $n=10$ genes, and they work well on data with $n=10/18/20$ genes. If we want the models to work on data with many more genes, we need to train the models with corresponding data. However, when the gene number $n$ is large, the number of different GRNs increases exponentially, and the number of training samples should also increase violently. The time cost for generating such data can be quite large. Even with the same amount of training samples, the training speed of the $\mathrm{TE}(k)$ model is proportional to $n^4$, since the input length of the transformer encoder layer is proportional to $n^2$, and it needs to calculate the attention scores between all pairs of input tokens. Therefore, scaling TRENDY for a much larger $n$ is extremely time-consuming.

In this paper, the training data are generated by Eq.~\ref{eqnl} that simulates the dynamics of mRNA count as a continuous-state process. This might be an oversimplification of gene regulation in reality, where the gene state and the protein count also should be considered \cite{holehouse2020stochastic}. Besides, the experimental gene expression data often suffer from incomplete measurement, where many mRNAs are not recorded, and there are many missing values. Therefore, when we train on perfectly measured data and test on data with missing values, the performance is not guaranteed. We can manually add measurement errors and missing values to the training data to solve this problem.

In our testing, we find that adding covariance matrices in the transformer (TRENDY and tGENIE3) is beneficial, since the covariance matrices may contain extra information besides that contained in the inferred GRN. 

{TRENDY uses covariance matrices instead of raw gene expression data as input. One advantage is that calculating covariance matrices by graphical lasso regularizes the data. One disadvantage is that covariance only concerns second-order moment, while information contained in higher-order moments is abandoned.}

\section{Methods}

\subsection{Training data}
To generate training data for our TRENDY method, we use Eq.~\ref{eqnl} to numerically simulate gene expression data with the Euler-Maruyama method \cite{kloeden1992stochastic} for $n=10$ genes. Every time we generate a random ground truth GRN $A_{\mathrm{true}}$, where each element has probability $0.1/0.8/0.1$ to be $-1/0/1$. Then we use Eq.~\ref{eqnl} with this random $A_{\mathrm{true}}$ to run the simulation from time $0.0$ to time $1.0$ with time step $0.01$ to obtain one trajectory, where the initial state at time $0$ is random. This is repeated 100 times to obtain 100 trajectories, similar to the situation where we measure the single-cell expression levels of $n$ genes for 100 cells \cite{qian2020counting}. We only record expression levels at time points $0.0,0.1,0.2,\ldots,1.0$. We repeat the simulation to obtain $1.01\times 10^5$ samples, where $10^5$ samples are for training, and $10^3$ samples are for validation (used in hyperparameter tuning and early stopping).

\subsection{Loss function}
We use AUROC and AUPRC to measure the difference between $A_\Z$ and the inferred $A_{\mathrm{pred}}$. See Section~S2 for details. Unfortunately, for fixed $A_\Z$ and arbitrary $A_{\mathrm{pred}}$, AUROC and AUPRC can only take finitely many values, and are not continuous functions of $A_{\mathrm{pred}}$. We cannot train a neural network with a non-differentiable loss function like AUROC or AUPRC. There have been some surrogate loss functions for AUROC or AUPRC \cite{qi2021stochastic,yuan2023libauc}. We directly use the mean square error loss function, which also performs well. The loss for comparing two GRNs does not count the diagonal elements, since they represent the autoregulation that cannot be inferred by the methods in this paper \cite{wang2023inference}.

\subsection{Segment embedding}
For $\mathrm{TE}(k)$ model with $k>1$ input matrices, after processing them with the linear embedding layer, we need to add segment embeddings to them, { so that the model can differentiate between different input matrices} \cite{kenton2019bert}. We give the segment ID $0/1/2/\ldots$ to the first/second/third/$\ldots$ group of processed inputs, and use an embedding layer to map each segment ID to a $d$-dimensional embedding vector. Repeat this embedding vector $n\times n$ times to obtain an $(n\times n\times d)$ array, and add it to the $(n\times n\times d)$ arrays after the linear embedding layer. { In our numerical simulations, we find that this segment embedding layer slightly improves the performance.}

\subsection{Positional encoding}
For the $\mathrm{TE}(k)$ model, the 2-dimensional positional encoding layer \cite{xu2023stgrns} adds an $(n\times n\times d)$ array $\text{PE}$ to each of the embedded input, so that the model knows that the input is two-dimensional. For $x$ and $y$ in $1,2,\ldots,n$ and $j$ in $1,2,\ldots,d/4$, $\text{PE}$ is defined as 
\begin{equation}
    \begin{split}
         &\text{PE}[x,y,2j-1]=\text{PE}[x,y,2j-1+d/2]=\sin[(y-1)\times 10^{-32(j-1)/d}],\\
&\text{PE}[x,y,2j]=\text{PE}[x,y,2j+d/2]=\cos[(y-1)\times 10^{-32(j-1)/d}],\\
& \text{PE}[x,y,2j-1+d/4]=\text{PE}[x,y,2j-1+3d/4]=\sin[(x-1)\times 10^{-32(j-1)/d}],\\
& \text{PE}[x,y,2j+d/4]=\text{PE}[x,y,2j+3d/4]=\cos[(x-1)\times 10^{-32(j-1)/d}].
    \end{split}
\end{equation}
In our numerical simulations, we find that this 2-dimensional positional encoding is crucial for the $\mathrm{TE}(k)$ model to produce better results.

\subsection{Training setting and cost}
All $\mathrm{TE}(k)$ models used in this paper are trained for 100 epochs. For each epoch, we evaluate the model performance on the validation data set. If the performance does not improve for 10 consecutive epochs, we stop training early. For all transformer encoder layers, the dropout rate is 0.1. The optimizer is Adam with learning rate 0.001.

Data generation and model training are conducted on a desktop computer with Intel i7-13700 CPU and NVIDIA RTX 4080 GPU. Data generation takes about one week with CPU parallelization. Training for each $\mathrm{TE}(k)$ model takes about two hours with GPU acceleration. After training, the time cost of applying TRENDY on a data set with $\sim 10$ genes and $\sim 100$ cells is under one second.

\appendix

\renewcommand{\thesection}{S\arabic{section}}
\renewcommand{\thefigure}{S\arabic{figure}}
\renewcommand{\thetable}{S\arabic{table}}

\setcounter{section}{0}
\setcounter{figure}{0}
\setcounter{table}{0}

\section{Details of WENDY method}
We start with a general SDE for gene expression levels $X_j(t)$:
\[\frac{\mathrm{d}X_i(t)}{\mathrm{d}t}=f(X_1,\ldots,X_n)+c_{1,i}-c_{2,i}X_i+\mathrm{noise}.\]
Here $f$ describes the interactions of different genes; $c_{1,i}$ is the synthesis rate; $c_{2,i}$ is the degradation rate.

After some simplifications and linearizations, we have 
\[\frac{\mathrm{d}X(t)}{\mathrm{d}t}=X(t)A+c+X(t)\odot\mathrm{d}\sigma W(t).\]
Here $X(t)=[X_1(t),\ldots,X_n(t)]$, $A$ is the GRN, $W(t)$ is an $n$-dimensional standard Brownian motion, and $\odot$ is the entrywise (Hadamard) product. 

Its approximated solution is
\[X(t)=X(0)(I+tA)+tc + X(0) \odot \epsilon(t),\]
where $\epsilon(t)=[\epsilon_1(t),\ldots,\epsilon_n(t)]$ is an $n$-dimensional normal random noise with $0$ mean and diagonal covariance matrix.

Then we can calculate the covariance matrix
\[K(t)=\mathbb{E}\{[X(t)^\T-\mathbb{E}X(t)^\T][X(t)-\mathbb{E}X(t)]\},\]
and the final equation is
\[K(t)=(I+tA^\mathrm{T})K(0)(I+tA)+D+E,\]
where $D$ is an unknown diagonal matrix that depends on the variance of $\epsilon$, and $E$ is the error introduced by linearizations.

WENDY method solves $A$ from the last equation by the BFGS algorithm.

\section{Calculation of AUROC and AUPRC}
In our setting, each edge in the true GRN can only take three values (labels): -1 (negative regulation), 0 (no regulation), and 1 (positive regulation). In the inferred GRN, each edge corresponds to a real number that can be mapped to those three labels. We need to evaluate the inferred GRN as a classification problem.

AUROC (Area Under the Receiver Operating Characteristic Curve) and AUPRC (Area Under the Precision-Recall Curve) are metrics for evaluating the performance of classification models (originally for binary classification). They are both between 0 and 1. If the case has perfect match, where the sorted predicted results and the sorted true labels have the same order, then AUROC and AUPRC are 1. If the order is fully reversed, AUROC and AUPRC are 0. Notice that only the order matters, and the predicted results do not need to match the true labels numerically. Although they have similar properties, AUROC is more valuable when different true labels are relatively balanced, and AUPRC is more informative with imbalanced labels. Therefore, we use both to evaluate the inferred GRN.

We use a toy example to illustrate the calculation of AUROC and AUPRC, given the ground truth GRN and the inferred GRN.

Assume that the ground truth GRN and the inferred GRN are
\[A_{\mathrm{true}} =
\begin{bmatrix}
  0 &  0 &  0 &  1 &  1 \\
  1 &  1 & -1 &  0 &  0 \\
  0 & -1 &  1 &  0 &  1 \\
  1 & -1 & -1 &  1 &  0 \\
  0 & -1 & -1 & -1 & -1
\end{bmatrix},\]
and \[A_{\mathrm{pred}} =
\begin{bmatrix}
  0.45 &  0.15 &  0.30 & -0.94 &  0.24 \\
  0.02 &  0.09 & -0.28 & -0.84 & -0.61 \\
  0.90 & -0.18 & -0.07 & -0.89 & -0.92 \\
  0.43 &  0.08 &  0.02 & -0.51 &  0.49 \\
 -0.66 & -0.62 & -0.08 & -0.43 & -0.51
\end{bmatrix}.\]

Since all GRN inference methods in the main text cannot handle autoregulation, we omit edges $(i, i)$. For each edge $(i,j)$ with $i\ne j$, we know the true label $A_{\mathrm{true}}[i,j]$ and the predicted value $A_{\mathrm{pred}}[i,j]$. Since the true label has three possibilities, but AUROC and AUPRC are originally defined for binary classification, we need to consider every two different labels.

Consider all edges $(i,j)$ with $i\ne j$ and true label $0$ or $1$. List $(A_{\mathrm{pred}}[i,j],A_{\mathrm{true}}[i,j])$ and sort by $A_{\mathrm{pred}}[i,j]$. Then we have
(-0.94, 1), (-0.92, 1), (-0.89, 0), (-0.84, 0), (-0.66, 0), (-0.61, 0), ( 0.02, 1), ( 0.15, 0), ( 0.24, 1), ( 0.30, 0), ( 0.43, 1), ( 0.49, 0), ( 0.90, 0).

For a given threshold $T$, we can count 

\noindent true positives (TP): $A_{\mathrm{true}}[i,j]=1$, $A_{\mathrm{pred}}[i,j]>T$;

\noindent true negatives (TN): $A_{\mathrm{true}}[i,j]=0$, $A_{\mathrm{pred}}[i,j]\le T$;

\noindent false positives (FP): $A_{\mathrm{true}}[i,j]=0$, $A_{\mathrm{pred}}[i,j]>T$;

\noindent false negatives (FN): $A_{\mathrm{true}}[i,j]=1$, $A_{\mathrm{pred}}[i,j]\le T$.

Then we can calculate 

\noindent true positive rate (TPR): TP/(TP+FN);

\noindent false positive rate (FPR): FP/(FP+TN).

For instance, if the threshold to be $0$, then 

\noindent TP = 3, \quad TN = 4, \quad FP = 4, \quad FN = 2, \quad\text{TPR} = 0.6, \quad \text{FPR} = 0.5.

\begin{figure}
    \centering
        \includegraphics[width=0.9\textwidth]{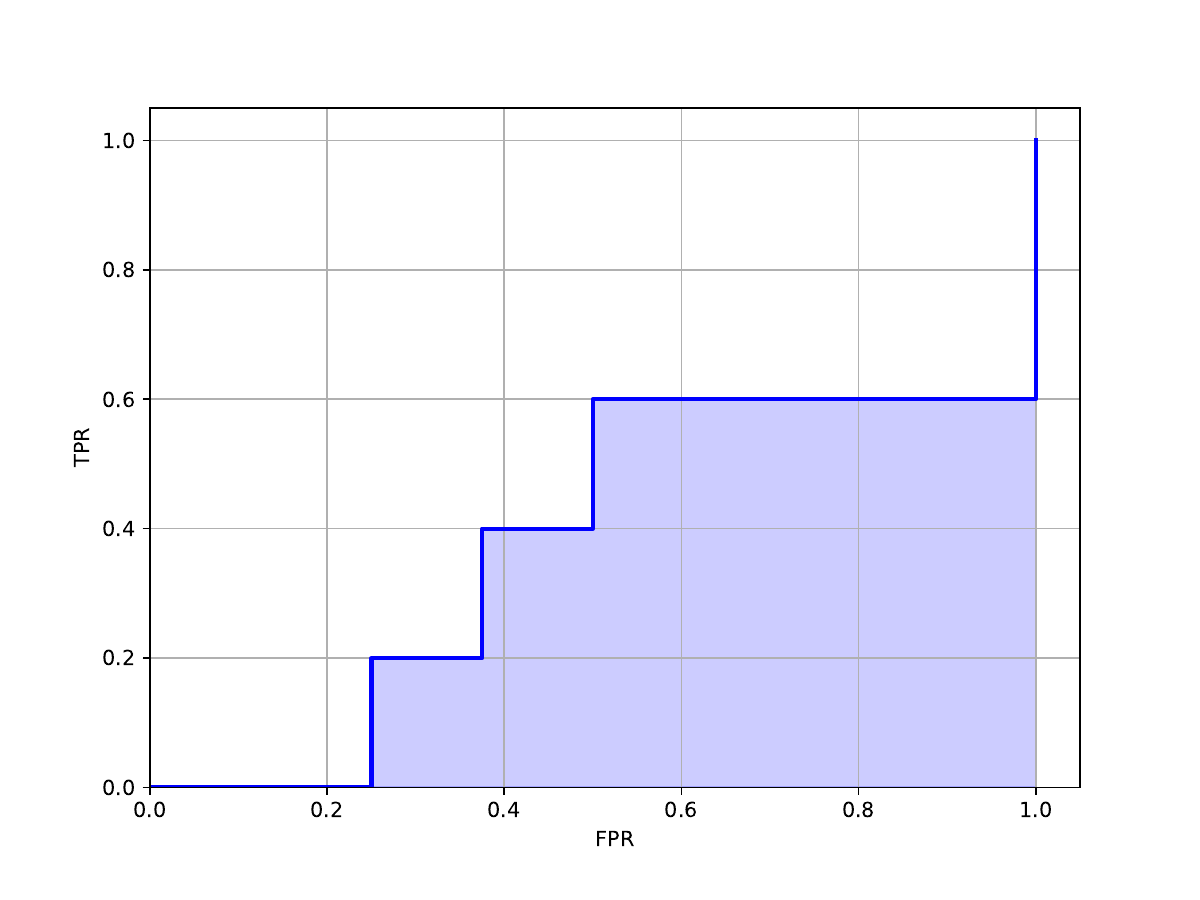}
    \caption{Example of ROC curve, where AUROC is the shaded area.}
    \label{roc}
\end{figure}

We let the threshold $T$ change from -1 to 1, and draw the corresponding points (FPR, TPR) in the square $[0,1]\times [0,1]$: (0.0, 0.0), (0.125, 0.0), (0.25, 0.0), (0.25, 0.2), (0.375, 0.2), (0.375, 0.4), (0.5, 0.4), (0.5, 0.6), (1.0, 0.6), (1.0, 1.0). Connect these points to obtain the receiver operating characteristic (ROC) curve, and AUROC is the area under this curve. See Fig.~\ref{roc} for the illustration, where AUROC is 0.375.

For true labels -1 and 1, repeat this procedure, and the AUROC[-1,1] is 0.571. For true labels -1 and 0, repeat this procedure, and the AUROC[-1,0] is 0.518. 

AUROC for two labels has an equivalent definition: in all pairs of edges $(i,j)$ with $i\ne j$ and $(p,q)$ with $p\ne q$, so that $A_{\mathrm{true}}[i,j]<A_{\mathrm{true}}[p,q]$, AUROC is the proportion that $A_{\mathrm{pred}}[i,j]<A_{\mathrm{pred}}[p,q]$.

Inspired by this definition, the AUROC for three labels should be the weighted average of three AUROC for two labels, where the weight is the number of pairs with different labels. In this example, there are 5 edges with true label 1, 8 edges with true label 0, and 7 edges with true label -1. Therefore, the final AUROC is 
\[\frac{5\times 8 \times \mathrm{AUROC}[0,1]+5\times 7 \times \mathrm{AUROC}[-1,1]+8\times 7 \times \mathrm{AUROC}[-1,0]}{5\times 8 +5\times 7 + 8\times 7}=0.489.\]

To calculate AUPRC, just replace TPR by Precision=TP/(TP+FP), and replace FPR by Recall=TP/(TP+FN).

\section{Training and testing details}

During training, for variants of WENDY, since they only need two time points, we fix the first time point to be $0.0$, and the second time point can be $0.1,0.2,\ldots,1.0$. Therefore, each simulated sample corresponds to 10 pairs of $K_0$ and $K_t$. This means that the training set has $10^6$ pairs of $K_0$ and $K_t$, where each $t$ in $0.1,0.2,\ldots,1.0$ corresponds to $10^5$ pairs. For variants of GENIE3, we consider each time point $t$ in $0.1,0.2,\ldots,1.0$ and calculate the inferred GRN $A_\mathrm{G}$. Then there are $10^6$ pairs of $A_\mathrm{G}$ and $K_t$ for training the $\mathrm{TE}(k=2)$ model. For variants of SINCERITIES and NonlinearODEs, we use the data at all time points $0.0,0.1,0.2,\ldots,1.0$ to infer the GRN $A_\mathrm{S}$ and $A_\mathrm{N}$. Therefore, each method has $10^5$ samples for training the $\mathrm{TE}(k=1)$ model.

For SINC data set, there are 1000 samples for each value of $\sigma=0.01/0.1/1$, and each sample with a random $A_\Z$ has 100 cells measured at time points $0.0, 0.1,\ldots,1.0$. For WENDY and its variants, we set the first time point to $0.0$ and vary the second time point from $0.1$ to $1.0$. For GENIE3 and its variants, we set the time point to any of $0.1,\ldots,1.0$. For SINCERITIES, NonlinearODEs, and their variants, we use all time points. 

For DREAM4 data set, we use the 10-gene time series data set. There are five $A_\Z$, each corresponding to the levels of 10 genes for five cells at 21 time points. For WENDY and its variants, we consider any pair of time points and take average. For GENIE3 and its variants, we consider any one time point and take average. For SINCERITIES, NonlinearODEs, and their variants, we use any consecutive 11 time points and take average. 

For the THP-1 data set, there are eight time points, each with 120 cells measured. We consider the expression levels of 20 genes. See Fig.~S2 for the ground truth GRN. For WENDY and its variants, we consider any pair of time points and take average. For GENIE3 and its variants, we consider any one time point and take average. For SINCERITIES, NonlinearODEs, and their variants, we use all eight time points. 

For the hESC data set, there are six time points, each with 66--172 cells measured. The same as in a previous paper \cite{wang2024gene}, we consider the expression levels of 18 genes. See Fig.~S3 for the ground truth GRN. For WENDY and its variants, we consider any pair of time points and take average. For GENIE3 and its variants, we consider any one time point and take average. For SINCERITIES, NonlinearODEs, and their variants, we use all six time points.

\section{Heat maps of true GRN for experimental data}
\begin{figure}
    \centering
        \includegraphics[width=0.9\textwidth]{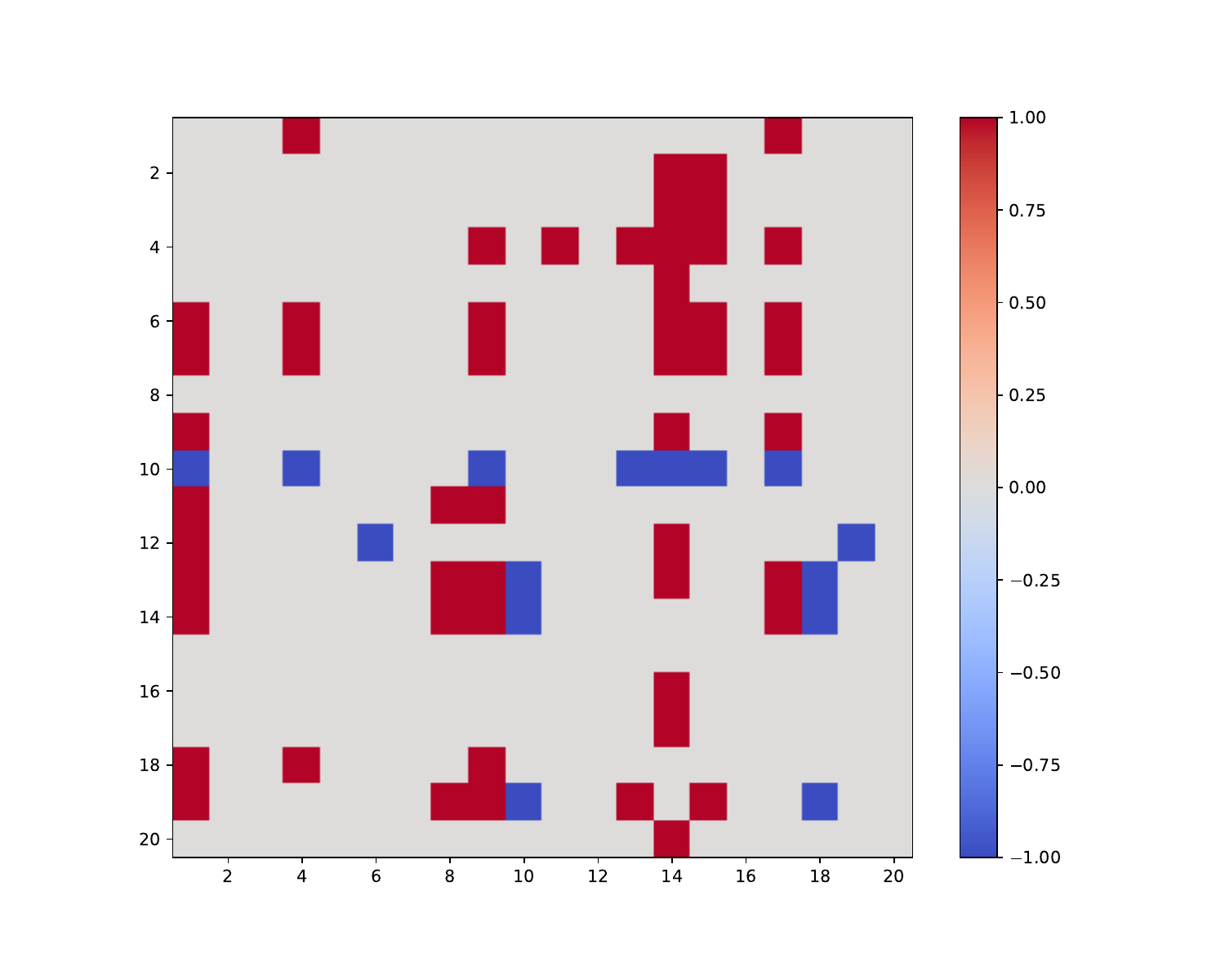}
    \caption{Heat map of the true GRN of THP-1 data set.}
    \label{t1a}
\end{figure}

See Fig.~\ref{t1a} for the heat map of the true GRN of THP-1 data set. Possible values are 1, -1, and 0, meaning positive regulation, negative regulation, and no regulation.

\begin{figure}
    \centering
        \includegraphics[width=0.9\textwidth]{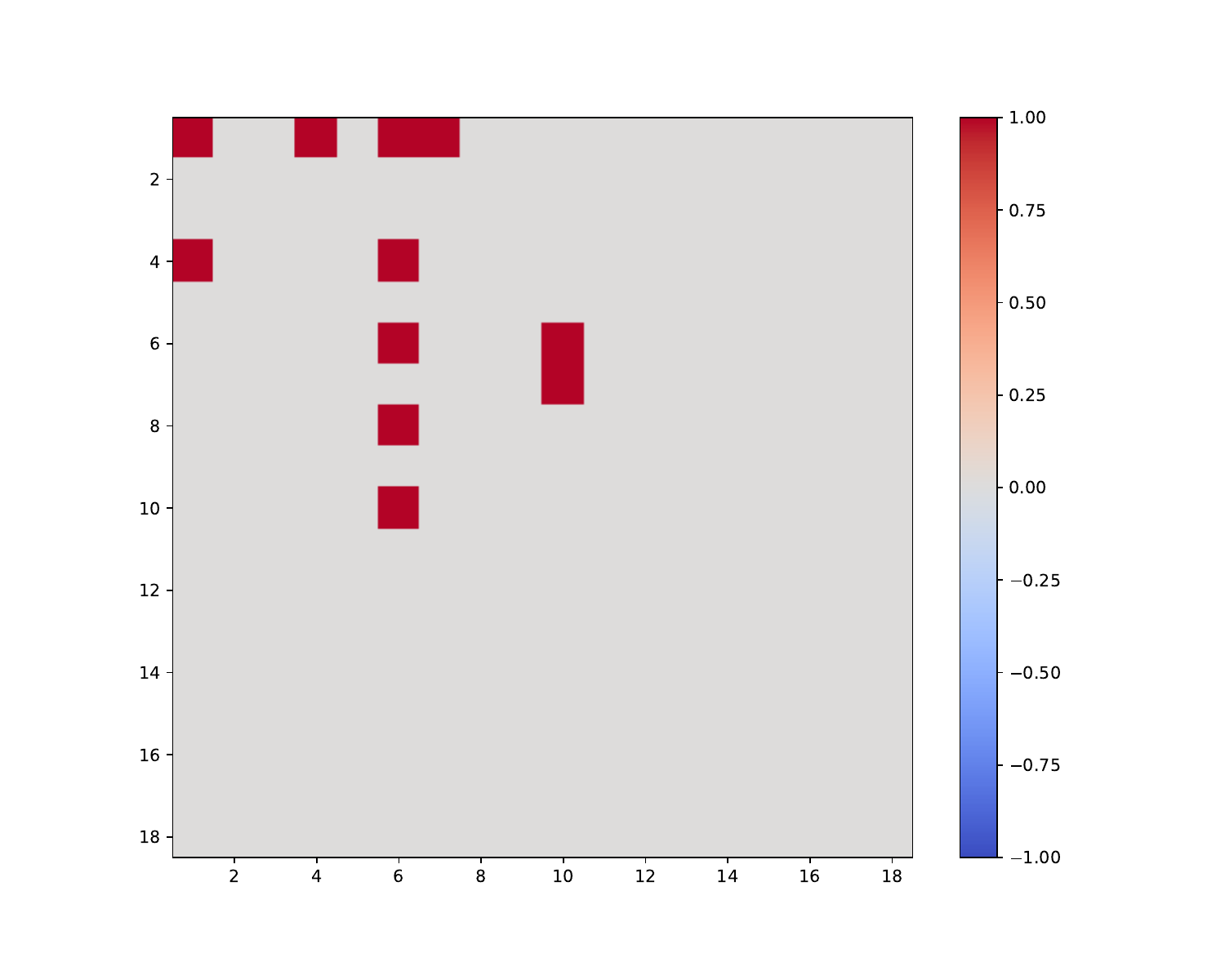}
    \caption{Heat map of the true GRN of hESC data set.}
    \label{ha}
\end{figure}

See Fig.~\ref{ha} for the heat map of the true GRN of hESC data set. Possible values are 1 and 0, meaning regulation and no regulation. Notice that almost all values are 0.

\section{Detailed performance measurements}

\begin{table}[ht]
\centering
\resizebox{\textwidth}{!}{%
\begin{tabular}{lccccccccccc}
\toprule
 & \multicolumn{2}{c}{$\sigma=0.01$} & & \multicolumn{2}{c}{$\sigma=0.1$} & & \multicolumn{2}{c}{$\sigma=1$} & & \multicolumn{2}{c}{Average} \\

{Method} & {AUROC} & {AUPRC} & & {AUROC} & {AUPRC} & & {AUROC} & {AUPRC} & & {AUROC} & {AUPRC} \\
\midrule
WENDY          & 0.6290 & 0.5897 & & 0.6654 & 0.6109 & & 0.6008 & 0.5782 & & 0.6317 & 0.5929 \\
TRENDY         & 0.7533 & 0.6908 & & 0.8547 & 0.7519 & & 0.6678 & 0.6087 & & 0.7586 & 0.6838 \\
nWENDY         & 0.5227 & 0.5431 & & 0.5131 & 0.5447 & & 0.5045 & 0.5314 & & 0.5134 & 0.5397 \\
bWENDY         & 0.5227 & 0.5413 & & 0.5129 & 0.5431 & & 0.5052 & 0.5312 & & 0.5136 & 0.5385 \\
&&&&&&&&&&&\\
GENIE3         & 0.4487 & 0.5378 & & 0.4062 & 0.5061 & & 0.3965 & 0.4881 & & 0.4171 & 0.5107 \\
tGENIE3        & 0.7844 & 0.7189 & & 0.8703 & 0.7672 & & 0.6757 & 0.6040 & & 0.7768 & 0.6967 \\
nGENIE3        & 0.4403 & 0.5241 & & 0.3985 & 0.4976 & & 0.3689 & 0.4748 & & 0.4026 & 0.4988 \\
bGENIE3        & 0.4418 & 0.5223 & & 0.3983 & 0.4959 & & 0.3693 & 0.4746 & & 0.4031 & 0.4976 \\
&&&&&&&&&&&\\
SINCERITIES    & 0.6493 & 0.5726 & & 0.6783 & 0.5829 & & 0.7154 & 0.5967 & & 0.6810 & 0.5841 \\
tSINCERITIES   & 0.7192 & 0.6185 & & 0.7964 & 0.6637 & & 0.7661 & 0.6294 & & 0.7606 & 0.6372 \\
nSINCERITIES   & 0.5401 & 0.5369 & & 0.5394 & 0.5387 & & 0.5332 & 0.5363 & & 0.5376 & 0.5373 \\
bSINCERITIES   & 0.5422 & 0.5375 & & 0.5450 & 0.5409 & & 0.5374 & 0.5381 & & 0.5415 & 0.5388 \\
&&&&&&&&&&&\\
NonlinearODEs  & 0.5026 & 0.5243 & & 0.5076 & 0.5313 & & 0.5068 & 0.5301 & & 0.5057 & 0.5286 \\
tNonlinearODEs & 0.5487 & 0.5460 & & 0.5976 & 0.5658 & & 0.5162 & 0.5333 & & 0.5542 & 0.5484 \\
nNonlinearODEs & 0.4971 & 0.5205 & & 0.5053 & 0.5251 & & 0.5073 & 0.5236 & & 0.5032 & 0.5231 \\
bNonlinearODEs & 0.4972 & 0.5202 & & 0.5058 & 0.5260 & & 0.5081 & 0.5248 & & 0.5037 & 0.5237 \\
\bottomrule
\end{tabular}%
}
\caption{AUROC and AUPRC scores of all 16 methods on SINC data set. We list the scores separately for data with $\sigma=0.01$, $\sigma=0.1$, $\sigma=1$, and the average scores.}
\label{sinc}
\end{table}

\begin{table}[ht]
\centering
\resizebox{\textwidth}{!}{%
\begin{tabular}{lccccc cccccc}
\toprule
 & \multicolumn{2}{c}{DREAM4} & & \multicolumn{2}{c}{THP-1} & & \multicolumn{2}{c}{hESC} \\

{Method} & {AUROC} & {AUPRC} & & {AUROC} & {AUPRC} & & {AUROC} & {AUPRC} \\
\midrule
WENDY          & 0.4899 & 0.2080 & & 0.5261 & 0.3972 & & 0.4997 & 0.0392 \\
TRENDY         & 0.5341 & 0.2177 & & 0.5557 & 0.3669 & & 0.5311 & 0.0376 \\
nWENDY         & 0.5417 & 0.2254 & & 0.6112 & 0.4203 & & 0.4971 & 0.0372 \\
bWENDY         & 0.5421 & 0.2231 & & 0.6106 & 0.4205 & & 0.5070 & 0.0402 \\
GENIE3         & 0.5636 & 0.2286 & & 0.4484 & 0.3546 & & 0.5913 & 0.0468 \\
tGENIE3        & 0.4589 & 0.1799 & & 0.5506 & 0.3781 & & 0.6008 & 0.0435 \\
nGENIE3        & 0.5632 & 0.2261 & & 0.4861 & 0.3642 & & 0.5744 & 0.0462 \\
bGENIE3        & 0.5741 & 0.2284 & & 0.4792 & 0.3623 & & 0.5767 & 0.0488 \\
SINCERITIES    & 0.4908 & 0.1919 & & 0.6261 & 0.3852 & & 0.4198 & 0.0261 \\
tSINCERITIES   & 0.4995 & 0.2034 & & 0.5251 & 0.3412 & & 0.4871 & 0.0294 \\
nSINCERITIES   & 0.4999 & 0.1856 & & 0.5956 & 0.3900 & & 0.1955 & 0.0199 \\
bSINCERITIES   & 0.5040 & 0.1846 & & 0.6067 & 0.3798 & & 0.1842 & 0.0196 \\
NonlinearODEs  & 0.4806 & 0.1705 & & 0.5338 & 0.3486 & & 0.5971 & 0.0534 \\
tNonlinearODEs & 0.5712 & 0.2452 & & 0.4808 & 0.3302 & & 0.6233 & 0.0641 \\
nNonlinearODEs & 0.4791 & 0.1772 & & 0.5521 & 0.3482 & & 0.6008 & 0.0466 \\
bNonlinearODEs & 0.4856 & 0.1666 & & 0.5544 & 0.3498 & & 0.6040 & 0.0633 \\
\bottomrule
\end{tabular}%
}
\caption{AUROC and AUPRC scores of all 16 methods on DREAM4, THP-1, and hESC data sets.}
\label{others}
\end{table}

\section*{Acknowledgments}
The authors would like to thank Dr. Zikun Wang for helpful comments. YW would like to thank Dr. Mingda Zhang, Dr. Lingxue Zhu, and Mr. Vincent Zhang for an intriguing discussion. YP would like to thank Irving Institute for Cancer Dynamics of Columbia University for hosting the Summer Research Program.

\section*{Data and code availability}
Main function of TRENDY method, including a tutorial and the training files, along with other data and code files used in this paper, can be found in 
\begin{verbatim}
https://github.com/YueWangMathbio/TRENDY
\end{verbatim}

\bibliographystyle{unsrt}
\bibliography{trendy}

\begin{thebibliography}{10}

\bibitem{wang2022chronic}
Zikun Wang, Samantha Lincoln, Andrew~D Nguyen, Wanhe Li, and Michael~W Young.
\newblock Chronic sleep loss disrupts rhythmic gene expression in drosophila.
\newblock {\em Frontiers in Physiology}, 13:1048751, 2022.

\bibitem{wang2020identification}
Zikun Wang.
\newblock {\em Identification of Gene Expression Changes in Sleep Mutants
  Associated With Reduced Longevity in Drosophila}.
\newblock {PhD thesis}, The Rockefeller University, 59--74, 2020.

\bibitem{wang2020biological}
Yue Wang, J{\'e}r{\'e}mie Kropp, and Nadya Morozova.
\newblock Biological notion of positional information/value in morphogenesis
  theory.
\newblock {\em International Journal of Developmental Biology},
  64(10-11-12):453--463, 2020.

\bibitem{cheng2024reconstruction}
Yu-Chen Cheng, Yun Zhang, Shubham Tripathi, BV~Harshavardhan, Mohit~Kumar
  Jolly, Geoffrey Schiebinger, Herbert Levine, Thomas~O McDonald, and Franziska
  Michor.
\newblock Reconstruction of single-cell lineage trajectories and identification
  of diversity in fates during the epithelial-to-mesenchymal transition.
\newblock {\em Proceedings of the National Academy of Sciences},
  121(32):e2406842121, 2024.

\bibitem{cheng2022ex}
Rene Yu-Hong Cheng, King~L Hung, Tingting Zhang, Claire~M Stoffers, Andee~R
  Ott, Emmaline~R Suchland, Nathan~D Camp, Iram~F Khan, Swati Singh, Ying-Jen
  Yang, et~al.
\newblock Ex vivo engineered human plasma cells exhibit robust protein
  secretion and long-term engraftment in vivo.
\newblock {\em Nature Communications}, 13(1):6110, 2022.

\bibitem{sha2024reconstructing}
Yutong Sha, Yuchi Qiu, Peijie Zhou, and Qing Nie.
\newblock Reconstructing growth and dynamic trajectories from single-cell
  transcriptomics data.
\newblock {\em Nature Machine Intelligence}, 6(1):25--39, 2024.

\bibitem{myasnikova2020gene}
Ekaterina Myasnikova and Alexander Spirov.
\newblock Gene regulatory networks in drosophila early embryonic development as
  a model for the study of the temporal identity of neuroblasts.
\newblock {\em Biosystems}, 197:104192, 2020.

\bibitem{mcdonald2023computational}
Thomas~O McDonald, Yu-Chen Cheng, Christopher Graser, Phillip~B Nicol, Daniel
  Temko, and Franziska Michor.
\newblock Computational approaches to modelling and optimizing cancer
  treatment.
\newblock {\em Nature Reviews Bioengineering}, 1(10):695--711, 2023.

\bibitem{cheng2023mathematical}
Yu-Chen Cheng, Shayna Stein, Agostina Nardone, Weihan Liu, Wen Ma, Gabriella
  Cohen, Cristina Guarducci, Thomas~O McDonald, Rinath Jeselsohn, and Franziska
  Michor.
\newblock Mathematical modeling identifies optimum palbociclib-fulvestrant dose
  administration schedules for the treatment of patients with estrogen
  receptor--positive breast cancer.
\newblock {\em Cancer Research Communications}, 3(11):2331--2344, 2023.

\bibitem{angelini2022model}
Erin Angelini, Yue Wang, Joseph~Xu Zhou, Hong Qian, and Sui Huang.
\newblock A model for the intrinsic limit of cancer therapy: Duality of
  treatment-induced cell death and treatment-induced stemness.
\newblock {\em PLOS Computational Biology}, 18(7):e1010319, 2022.

\bibitem{arshad2014using}
Osama~A Arshad, Priyadharshini~S Venkatasubramani, Aniruddha Datta, and
  Jijayanagaram Venkatraj.
\newblock Using boolean logic modeling of gene regulatory networks to exploit
  the links between cancer and metabolism for therapeutic purposes.
\newblock {\em IEEE journal of biomedical and health informatics},
  20(1):399--407, 2014.

\bibitem{li2021chronic}
Wanhe Li, Zikun Wang, Sheyum Syed, Cheng Lyu, Samantha Lincoln, Jenna O’Neil,
  Andrew~D Nguyen, Irena Feng, and Michael~W Young.
\newblock Chronic social isolation signals starvation and reduces sleep in
  drosophila.
\newblock {\em Nature}, 597(7875):239--244, 2021.

\bibitem{vijayan2022internal}
Vikram Vijayan, Zikun Wang, Vikram Chandra, Arun Chakravorty, Rufei Li,
  Stephanie~L Sarbanes, Hessameddin Akhlaghpour, and Gaby Maimon.
\newblock An internal expectation guides drosophila egg-laying decisions.
\newblock {\em Science Advances}, 8(43):eabn3852, 2022.

\bibitem{axelrod2023drosophila}
Sofia Axelrod, Xiaoling Li, Yingwo Sun, Samantha Lincoln, Andrea Terceros,
  Jenna O’Neil, Zikun Wang, Andrew Nguyen, Aabha Vora, Carmen Spicer, et~al.
\newblock The drosophila blood--brain barrier regulates sleep via moody g
  protein-coupled receptor signaling.
\newblock {\em Proceedings of the National Academy of Sciences},
  120(42):e2309331120, 2023.

\bibitem{wang2024gene}
Yue Wang, Peng Zheng, Yu-Chen Cheng, Zikun Wang, and Aleksandr Aravkin.
\newblock Wendy: Covariance dynamics based gene regulatory network inference.
\newblock {\em Mathematical Biosciences}, 377:109284, 2024.

\bibitem{ma2020inference}
Baoshan Ma, Mingkun Fang, and Xiangtian Jiao.
\newblock Inference of gene regulatory networks based on nonlinear ordinary
  differential equations.
\newblock {\em Bioinformatics}, 36(19):4885--4893, 2020.

\bibitem{burdziak2023sckinetics}
Cassandra Burdziak, Chujun~Julia Zhao, Doron Haviv, Direna Alonso-Curbelo,
  Scott~W Lowe, and Dana Pe’er.
\newblock sckinetics: inference of regulatory velocity with single-cell
  transcriptomics data.
\newblock {\em Bioinformatics}, 39(Supplement\_1):i394--i403, 2023.

\bibitem{wang2023dictys}
Lingfei Wang, Nikolaos Trasanidis, Ting Wu, Guanlan Dong, Michael Hu, Daniel~E
  Bauer, and Luca Pinello.
\newblock Dictys: dynamic gene regulatory network dissects developmental
  continuum with single-cell multiomics.
\newblock {\em Nature Methods}, 20(9):1368--1378, 2023.

\bibitem{huynh2010inferring}
V{\^a}n~Anh Huynh-Thu, Alexandre Irrthum, Louis Wehenkel, and Pierre Geurts.
\newblock Inferring regulatory networks from expression data using tree-based
  methods.
\newblock {\em PloS one}, 5(9):e12776, 2010.

\bibitem{papili2018sincerities}
Nan Papili~Gao, SM~Minhaz Ud-Dean, Olivier Gandrillon, and Rudiyanto Gunawan.
\newblock {SINCERITIES}: inferring gene regulatory networks from time-stamped
  single cell transcriptional expression profiles.
\newblock {\em Bioinformatics}, 34(2):258--266, 2018.

\bibitem{zheng2019bixgboost}
Ruiqing Zheng, Min Li, Xiang Chen, Fang-Xiang Wu, Yi~Pan, and Jianxin Wang.
\newblock Bixgboost: a scalable, flexible boosting-based method for
  reconstructing gene regulatory networks.
\newblock {\em Bioinformatics}, 35(11):1893--1900, 2019.

\bibitem{huynh2018dyngenie3}
V{\^a}n~Anh Huynh-Thu and Pierre Geurts.
\newblock dyngenie3: dynamical genie3 for the inference of gene networks from
  time series expression data.
\newblock {\em Scientific reports}, 8(1):3384, 2018.

\bibitem{nauta2019causal}
Meike Nauta, Doina Bucur, and Christin Seifert.
\newblock Causal discovery with attention-based convolutional neural networks.
\newblock {\em Machine Learning and Knowledge Extraction}, 1(1):19, 2019.

\bibitem{kentzoglanakis2011swarm}
Kyriakos Kentzoglanakis and Matthew Poole.
\newblock A swarm intelligence framework for reconstructing gene networks:
  searching for biologically plausible architectures.
\newblock {\em IEEE/ACM Transactions on Computational Biology and
  Bioinformatics}, 9(2):358--371, 2011.

\bibitem{shu2021modeling}
Hantao Shu, Jingtian Zhou, Qiuyu Lian, Han Li, Dan Zhao, Jianyang Zeng, and
  Jianzhu Ma.
\newblock Modeling gene regulatory networks using neural network architectures.
\newblock {\em Nature Computational Science}, 1(7):491--501, 2021.

\bibitem{feng2023gene}
Ke~Feng, Hongyang Jiang, Chaoyi Yin, and Huiyan Sun.
\newblock Gene regulatory network inference based on causal discovery
  integrating with graph neural network.
\newblock {\em Quantitative Biology}, 11(4):434--450, 2023.

\bibitem{mao2023predicting}
Guo Mao, Zhengbin Pang, Ke~Zuo, Qinglin Wang, Xiangdong Pei, Xinhai Chen, and
  Jie Liu.
\newblock Predicting gene regulatory links from single-cell rna-seq data using
  graph neural networks.
\newblock {\em Briefings in Bioinformatics}, 24(6):bbad414, 2023.

\bibitem{vaswani2017attention}
Ashish Vaswani, Noam Shazeer, Niki Parmar, Jakob Uszkoreit, Llion Jones,
  Aidan~N Gomez, {\L}ukasz Kaiser, and Illia Polosukhin.
\newblock Attention is all you need.
\newblock In {\em Advances in Neural Information Processing Systems},
  volume~30, 2017.

\bibitem{xu2023stgrns}
Jing Xu, Aidi Zhang, Fang Liu, and Xiujun Zhang.
\newblock Stgrns: an interpretable transformer-based method for inferring gene
  regulatory networks from single-cell transcriptomic data.
\newblock {\em Bioinformatics}, 39(4):btad165, 2023.

\bibitem{shu2022boosting}
Hantao Shu, Fan Ding, Jingtian Zhou, Yexiang Xue, Dan Zhao, Jianyang Zeng, and
  Jianzhu Ma.
\newblock Boosting single-cell gene regulatory network reconstruction via
  bulk-cell transcriptomic data.
\newblock {\em Briefings in Bioinformatics}, 23(5):bbac389, 2022.

\bibitem{chan2017gene}
Thalia~E Chan, Michael~PH Stumpf, and Ann~C Babtie.
\newblock Gene regulatory network inference from single-cell data using
  multivariate information measures.
\newblock {\em Cell systems}, 5(3):251--267, 2017.

\bibitem{feizi2013network}
Soheil Feizi, Daniel Marbach, Muriel M{\'e}dard, and Manolis Kellis.
\newblock Network deconvolution as a general method to distinguish direct
  dependencies in networks.
\newblock {\em Nature biotechnology}, 31(8):726--733, 2013.

\bibitem{cao2013going}
Mengfei Cao, Hao Zhang, Jisoo Park, Noah~M Daniels, Mark~E Crovella, Lenore~J
  Cowen, and Benjamin Hescott.
\newblock Going the distance for protein function prediction: a new distance
  metric for protein interaction networks.
\newblock {\em PloS one}, 8(10):e76339, 2013.

\bibitem{pirayre2015brane}
Aur{\'e}lie Pirayre, Camille Couprie, Fr{\'e}d{\'e}rique Bidard, Laurent Duval,
  and Jean-Christophe Pesquet.
\newblock Brane cut: biologically-related a priori network enhancement with
  graph cuts for gene regulatory network inference.
\newblock {\em BMC bioinformatics}, 16:1--12, 2015.

\bibitem{pirayre2017brane}
Aur{\'e}lie Pirayre, Camille Couprie, Laurent Duval, and Jean-Christophe
  Pesquet.
\newblock Brane clust: Cluster-assisted gene regulatory network inference
  refinement.
\newblock {\em IEEE/ACM Transactions on Computational Biology and
  Bioinformatics}, 15(3):850--860, 2017.

\bibitem{wang2018network}
Bo~Wang, Armin Pourshafeie, Marinka Zitnik, Junjie Zhu, Carlos~D Bustamante,
  Serafim Batzoglou, and Jure Leskovec.
\newblock Network enhancement as a general method to denoise weighted
  biological networks.
\newblock {\em Nature communications}, 9(1):3108, 2018.

\bibitem{hache2009genge}
Hendrik Hache, Christoph Wierling, Hans Lehrach, and Ralf Herwig.
\newblock Genge: systematic generation of gene regulatory networks.
\newblock {\em Bioinformatics}, 25(9):1205--1207, 2009.

\bibitem{kamimoto2023dissecting}
Kenji Kamimoto, Blerta Stringa, Christy~M Hoffmann, Kunal Jindal, Lilianna
  Solnica-Krezel, and Samantha~A Morris.
\newblock Dissecting cell identity via network inference and in silico gene
  perturbation.
\newblock {\em Nature}, 614(7949):742--751, 2023.

\bibitem{dibaeinia2020sergio}
Payam Dibaeinia and Saurabh Sinha.
\newblock {SERGIO}: a single-cell expression simulator guided by gene
  regulatory networks.
\newblock {\em Cell systems}, 11(3):252--271, 2020.

\bibitem{schaffter2011genenetweaver}
Thomas Schaffter, Daniel Marbach, and Dario Floreano.
\newblock {GeneNetWeaver}: in silico benchmark generation and performance
  profiling of network inference methods.
\newblock {\em Bioinformatics}, 27(16):2263--2270, 2011.

\bibitem{tripathi2017sgnesr}
Shailesh Tripathi, Jason Lloyd-Price, Andre Ribeiro, Olli Yli-Harja, Matthias
  Dehmer, and Frank Emmert-Streib.
\newblock sgnesr: An r package for simulating gene expression data from an
  underlying real gene network structure considering delay parameters.
\newblock {\em BMC bioinformatics}, 18:1--12, 2017.

\bibitem{marouf2020realistic}
Mohamed Marouf, Pierre Machart, Vikas Bansal, Christoph Kilian, Daniel~S
  Magruder, Christian~F Krebs, and Stefan Bonn.
\newblock Realistic in silico generation and augmentation of single-cell
  rna-seq data using generative adversarial networks.
\newblock {\em Nature communications}, 11(1):166, 2020.

\bibitem{pinna2010knockouts}
Andrea Pinna, Nicola Soranzo, and Alberto De~La~Fuente.
\newblock From knockouts to networks: establishing direct cause-effect
  relationships through graph analysis.
\newblock {\em PloS one}, 5(10):e12912, 2010.

\bibitem{wang2022inference}
Yue Wang and Zikun Wang.
\newblock Inference on the structure of gene regulatory networks.
\newblock {\em Journal of Theoretical Biology}, 539:111055, 2022.

\bibitem{friedman2008sparse}
Jerome Friedman, Trevor Hastie, and Robert Tibshirani.
\newblock Sparse inverse covariance estimation with the graphical lasso.
\newblock {\em Biostatistics}, 9(3):432--441, 2008.

\bibitem{ramos2019physical}
Alexandre~F Ramos and John Reinitz.
\newblock Physical implications of so (2, 1) symmetry in exact solutions for a
  self-repressing gene.
\newblock {\em The Journal of Chemical Physics}, 151(4), 2019.

\bibitem{giovanini2020comparative}
Guilherme Giovanini, Alan~U Sabino, Luciana~RC Barros, Alexandre~F Ramos,
  et~al.
\newblock A comparative analysis of noise properties of stochastic binary
  models for a self-repressing and for an externally regulating gene.
\newblock {\em Mathematical Biosciences and Engineering}, 2020.

\bibitem{holehouse2020stochastic}
James Holehouse, Zhixing Cao, and Ramon Grima.
\newblock Stochastic modeling of autoregulatory genetic feedback loops: A
  review and comparative study.
\newblock {\em Biophysical Journal}, 118(7):1517--1525, 2020.

\bibitem{pare2009visualization}
Adam Par{\'e}, Derek Lemons, Dave Kosman, William Beaver, Yoav Freund, and
  William McGinnis.
\newblock Visualization of individual scr mrnas during drosophila embryogenesis
  yields evidence for transcriptional bursting.
\newblock {\em Current biology}, 19(23):2037--2042, 2009.

\bibitem{bothma2014dynamic}
Jacques~P Bothma, Hernan~G Garcia, Emilia Esposito, Gavin Schlissel, Thomas
  Gregor, and Michael Levine.
\newblock Dynamic regulation of eve stripe 2 expression reveals transcriptional
  bursts in living drosophila embryos.
\newblock {\em Proceedings of the National Academy of Sciences},
  111(29):10598--10603, 2014.

\bibitem{leyes2023transcriptional}
Emilia~A Leyes~Porello, Robert~T Trudeau, and Bomyi Lim.
\newblock Transcriptional bursting: stochasticity in deterministic development.
\newblock {\em Development}, 150(12):dev201546, 2023.

\bibitem{marbach2012wisdom}
Daniel Marbach, James~C Costello, Robert K{\"u}ffner, Nicole~M Vega, Robert~J
  Prill, Diogo~M Camacho, Kyle~R Allison, Manolis Kellis, James~J Collins,
  et~al.
\newblock Wisdom of crowds for robust gene network inference.
\newblock {\em Nature methods}, 9(8):796--804, 2012.

\bibitem{kouno2013temporal}
Tsukasa Kouno, Michiel de~Hoon, Jessica~C Mar, Yasuhiro Tomaru, Mitsuoki
  Kawano, Piero Carninci, Harukazu Suzuki, Yoshihide Hayashizaki, and Jay~W
  Shin.
\newblock Temporal dynamics and transcriptional control using single-cell gene
  expression analysis.
\newblock {\em Genome biology}, 14:1--12, 2013.

\bibitem{chu2016single}
Li-Fang Chu, Ning Leng, Jue Zhang, Zhonggang Hou, Daniel Mamott, David~T
  Vereide, Jeea Choi, Christina Kendziorski, Ron Stewart, and James~A Thomson.
\newblock Single-cell rna-seq reveals novel regulators of human embryonic stem
  cell differentiation to definitive endoderm.
\newblock {\em Genome biology}, 17:1--20, 2016.

\bibitem{liu2020fully}
Yi~Liu, Kenneth Barr, and John Reinitz.
\newblock Fully interpretable deep learning model of transcriptional control.
\newblock {\em Bioinformatics}, 36(Supplement\_1):i499--i507, 2020.

\bibitem{kloeden1992stochastic}
Peter~E Kloeden, Eckhard Platen, Peter~E Kloeden, and Eckhard Platen.
\newblock {\em Stochastic differential equations}.
\newblock Springer, 1992.

\bibitem{qian2020counting}
Hong Qian and Yu-Chen Cheng.
\newblock Counting single cells and computing their heterogeneity: from
  phenotypic frequencies to mean value of a quantitative biomarker.
\newblock {\em Quantitative Biology}, 8(2):172--176, 2020.

\bibitem{qi2021stochastic}
Qi~Qi, Youzhi Luo, Zhao Xu, Shuiwang Ji, and Tianbao Yang.
\newblock Stochastic optimization of areas under precision-recall curves with
  provable convergence.
\newblock {\em Advances in neural information processing systems},
  34:1752--1765, 2021.

\bibitem{yuan2023libauc}
Zhuoning Yuan, Dixian Zhu, Zi-Hao Qiu, Gang Li, Xuanhui Wang, and Tianbao Yang.
\newblock Libauc: A deep learning library for x-risk optimization.
\newblock In {\em Proceedings of the 29th ACM SIGKDD Conference on Knowledge
  Discovery and Data Mining}, pages 5487--5499, 2023.

\bibitem{wang2023inference}
Yue Wang and Siqi He.
\newblock Inference on autoregulation in gene expression with variance-to-mean
  ratio.
\newblock {\em Journal of Mathematical Biology}, 86(5):87, 2023.

\bibitem{kenton2019bert}
Jacob Devlin Ming-Wei~Chang Kenton and Lee~Kristina Toutanova.
\newblock Bert: Pre-training of deep bidirectional transformers for language
  understanding.
\newblock In {\em Proceedings of naacL-HLT}, volume~1, page~2. Minneapolis,
  Minnesota, 2019.

\end{thebibliography}
\end{document}